\documentclass[prb,
superscriptaddress,showpacs,amsmath,amssymb]{revtex4}
\usepackage{amsfonts}
\usepackage{bm}
\usepackage{verbatim}

\begin{document}

\title{From Landau two-fluid model to de Sitter Universe}

\author{G.E. Volovik}

\affiliation{ L.~D.~Landau Institute for
Theoretical Physics, 117940 Moscow, Russia}

\date{\today}

\begin{abstract}
{Condensed matter analogs are useful when considering phenomena associated with the quantum vacuum. This is because in condensed matter, we know physics both in the infrared and in the ultraviolet limits, whereas in particle physics and gravity, physics on the trans-Planckian scale is unknown. One of the cornerstones of the connection between non-relativistic condensed matter and modern relativistic theories is the two-fluid hydrodynamics of superfluid helium, developed by Landau and Khalatnikov (Isaak Markovich Khalatnikov later founded and headed the Landau Institute). 
The dynamics and thermodynamics of the de Sitter state of the expansion of the Universe bear some features of a multi-fluid system. There are actually three components: the quantum vacuum, the gravitational component and relativistic matter.
The expanding de Sitter vacuum serves as a thermal bath with local temperature, which is twice the Gibbons-Hawking temperature associated with the temperature of the cosmological horizon. This local temperature leads to the heating of the matter component and the gravitational component, which behaves as Zel'dovich stiff matter and represents dark matter.  In equilibrium and in the absence of ordinary matter, the positive partial pressure of the dark matter compensates the negative partial pressure of the quantum vacuum. That is why, in full equilibrium, the total pressure is zero.  This is rather similar to the correspondingly superfluid and normal components of superfluid liquid, which together produce a zero pressure of the liquid in the absence of environment. We propose the phenomenological theory, which describes the dynamics of dark energy and dark matter. If one assumes that, in the dynamics, gravitational dark matter behaves like real Zel'dovich stiff matter, it is shown that both components undergo power law decay due to the energy exchange between these components. It then follows that their values at the present time are of the correct order of magnitude. We also consider other problems through the prism of condensed matter physics, including black holes and the Planck constant.

}
\end{abstract}

\maketitle

\newcommand{\nh}{\slash\!\!\!h}
 \tableofcontents


\section{Introduction}
\label{Introduction}

The de Sitter universe has no ordinary matter content and is entirely supported by a quantum vacuum (called dark energy). This state is thought to have been in the early stages of inflation, and is now being approached again. Being isotropic and homogeneous in space and time, the de Sitter state is unique in cosmology. Isotropy and homogeneity allow us to construct a thermodynamics for this state that is in many ways similar to the thermodynamics of isotropic and homogeneous condensed matter.

The vacuum of the de Sitter spacetime is characterized  by the local temperature  $T=H/\pi$, where $H$ is the Hubble parameter, see Ref. \cite{Volovik2023k,Volovik2024} and references therein. This temperature describes the thermal processes of decay of the composite particles and the other activation processes, which are energetically forbidden in the Minkowski spacetime, but are allowed in the de Sitter background, see also Refs. \cite{Maldacena2015,Reece2023,Maxfield2022}. In particular, this temperature determines the probability of the ionization of an atom in the de Sitter environment, $\exp(-E/T)$, where $E$ is the ionization potential. This activation temperature is twice the Gibbons-Hawking\cite{GH1977} temperature $T_{\rm GH}=H/2\pi$ of the cosmological horizon, $T=2T_{\rm GH}$. 
As distinct from the $T_{\rm GH}$, the activation temperature has no relation to the cosmological horizon. It describes the local processes, which take place well inside the cosmological horizon. 

The factor of 2 difference between the local temperature and the temperature of the Hawking radiation from the cosmological horizon measured by the same observer has simple explanation. In case of the Hawking radiation the observer detects particles radiated from the horizon but cannot detect their partners, that are simultaneously (coherently) created on the other side of the horizon. This is similar to the doubling of the Hawking temperature of black holes suggested by  't Hooft.\cite{Hooft2022,Hooft2023,Hooft2024} In case of the black hole horizon, the suggested partner is in the mirror image of the black hole spacetime -- "quantum clone inside black hole".
The factor of 2 difference between the two temperatures is also supported by topological arguments.\cite{Kusmartsev2025,Volovik2025f}

The local temperature $T=H/\pi$ determines the local temperature in de Sitter spacetime, giving rise to the local entropy and the modified Gibbs-Duhem relation of the de Sitter state. The latter is modified due to the gravitational degrees of freedom: the gravitational coupling $K=1/16\pi G$  and the scalar Riemann curvature ${\cal R}$. These are the thermodynamically conjugate variables, which are similar to chemical potential and particle density in condensed matter. As a result, the de Sitter state can be considered as a mixture of two components: the vacuum energy (dark energy) and the gravitational component, which behaves as Zel'dovich stiff matter and has some analogy with dark matter. It is this dark matter component, which is responsible for the thermodynamics of the de Sitter state with equilibrium temperature $T=H/\pi$.
Together with the ordinary matter (which includes the real dark matter), the expanding Friedmann-Robertson-Walker Universe is represented by three components: dark energy, gravitational dark matter and ordinary matter. 

This is the analogue of the  Landau-Khalatnikov two-fluid hydrodynamics, which we consider in Section \ref{LandauKhalatnikov}.
This hydrodynamic theory became the source of different connections between the physics of condensed matter and relativistic physics, which includes quantum field theories, quantum gravity, and physics of the quantum vacuum.
The natural consequence of the two-fluid hydrodynamics is the acoustic metric in Sec.\ref{RiverModelSec}, which is produced by the "flow of the vacuum" -- the superfluid motion of the liquid. The excitations of the "superfluid vacuum" -- quasiparticles -- form analogue of matter on the background of the moving "quantum vacuum".
There are different directions in the extension of the analogue gravity. Gravity emerges in the systems, which have the point nodes in the fermionic spectrum, such as the chiral superfluid $^3$He-A and Weyl semimetals, see Sec. \ref{WeylPointSec}. Another scenario of formation of gravity comes from the elasticity tetrads, which describe deformations of crystals in Sec. \ref{SuperlasticSec}. In this case the quantum vacuum is considered as a superplastic crystal. Superfluid $^3$He-B suggests scenario, in which the gravitational tetrads emerge as the bilinear combination of fermions fields, see Sec.\ref{DIakonovDec}

In Section \ref{HeatBathSec}  we consider the thermodynamics of the de Sitter vacuum. We show that matter perceives the de Sitter state of the quantum vacuum as thermal bath with temperature $T=\hbar H/\pi$. This is demonstrated in Sec. \ref{AtomInDS} on example of ionization of the hydrogen atom in de Sitter environment, and in Sec.\ref{ProtonSec} on example of proton decay. In both cases the temperature $T=\hbar H/\pi$ determines the activation process. This local temperature determines the local thermodynamics of the de Sitter state: the energy density, free energy and entropy density in Sec.\ref{LocalT}. In Sec.\ref{holography} the holographic bulk-surface connection is obtained between the total entropy of the Hubble volume and the Gibbons-Hawking surface entropy of the cosmological horizon.

In Section \ref{dSthermodynamicsSec} the two-component thermodynamics of de Sitter state is discussed. This state can be represented in terms dark energy (vacuum) and dark matter (gravitational degrees of freedom). The latter are represented by the pair of the thermodynamically
conjugate variables: the gravitational coupling $K=1/16\pi G$
and the Riemann curvature ${\cal R}$. The equation of state of this dark matter is the same as that of Zeldovich stiff matter in Sec. \ref{StiffMatterSec} and of Landau Fermi liquid in Sec. \ref{LandauFermiSec}.

 In Section \ref{dSdynamicsSec} we consider the dynamics of de Sitter state which may follow from its thermodynamics. In the suggested phenomenological approach, it is assumed that the equations are modified in such a way, that away from equilibrium, the gravitational dark matter behaves as the real stiff matter. The latter experiences the cool-down due to expansion, which is partially compensated by the energy exchange with the dark energy component. The energy exchange leads to the loss of the dark energy, and as a result both dark components have the power-law decay. Both components reach at present time the correct order of magnitude. How this phenomenology is supported by microscopic theory is an open question.
 
In Section \ref{MacroscopicTunnelSec} it is shown how the quantum tunneling processes determine the thermodynamics of black hole.
The quantum tunneling processes of  Hawking radiation of particles determine the temperature of the black hole. The macroscopic quantum tunneling, which describes processes of splitting of the black hole into the smaller parts, determines the black hole entropy. This entropy is non-extensive and is described by the Thallis-Cirto entropy in Sec. \ref{NonextensiveSec}.
 
 In Section \ref{UnruhSchwingerSec} we discuss the temperature of the Unruh effect and its relation to  the Schwinger pair production. The combined process which includes the creation of charged particles by electric field and then their acceleration by this electric field demonstrates that the Unruh temperature has the extra factor 2, which is in agreement with the analogue of Unruh effect in superfluids.

In Section \ref{CosmologicalProblem} the cosmological constant problems are discussed using the experience with superfluid liquids. The main lesson from the ground states of these liquids is that the fully equilibrium quantum vacuum has exactly zero vacuum energy. This follows from the thermodynamics, which is valid both for many-body condensed matter systems and for relativistic quantum vacuum. This solves the main cosmological constant without any fine-tuning. 
The other cosmological problem is that it is not easy to reach the final equilibrium state. The de Sitter state has high symmetry, due to which it serves as the attractors in the dynamics of the Universe, and the final Minkowski vacuum can be obtained only by fine-tuning. However it is shown that the de Sitter vacuum is unstable. According to Section  \ref{HeatBathSec}, the de Sitter vacuum represents the heat bath for matter. This results in the thermal radiation of matter, which violates the de Sitter symmetry and thus promotes the relaxation to the Minkowski state with zero cosmological constant.

In Section \ref{PlanckConstantsSec} we discuss the dynamic origin of Planck constants $\hbar$ and $\nh =\hbar c$ as elements of the tetrads in Minkowski spacetime. We also discuss the space-time dimensions of the physical quantities. In particular the Planck constants $\hbar$ and $\nh$ have dimensions of time and space correspondingly, and thus they cannot be considered as the fundamental constants. The effective Planck constant $\nh_{\rm ac}$ emerging in the acoustic gravity in superfluid $^4$He has dimension of length and the size of the interatomic distance. Some dimensionless quantities can be quantized, which is considered on example of the entropies of the cosmological and black hole horizons. 

Among the authors of the cited papers are the members of Landau Institute:
Abrikosov, Beneslavskii, Berezinskii, Dzyaloshinsky, Feigel'man, Finkel'shtein, Gehshkenbein, Iordansky, Kamenshchik, Khalatnikov, Kopnin, Larkin, Polyakov, Rashba, Starobinsky.

\section{Landau-Khalatnikov two-fluid hydrodynamics and analogue gravity}
\label{LandauKhalatnikov}

 \subsection{River model of quantum vacuum}
\label{RiverModelSec}

Two-fluid hydrodynamics of superfluid $^4$He developed by Landau and Khalatnikov (see the Khalatnikov book \cite{KhalatnikovBook}) gives us important hints for understanding the general relativity and dynamics of quantum vacuum.
The relativistic character of the two-fluid hydrodynamics is manifested at low temperature, where the normal component of the liquid moving with velocity ${\bf v}_n$ is formed by the relativistic-like excitations with linear spectrum -- phonons. The coherent part of the liquid, which moves with velocity 
${\bf v}_s$, represents the "quantum vacuum", which provides the curved spacetime for these quasiparticles. This analogy can be seen already in Eq.(3.13) of the Khalatnikov book which corresponds to the Tolman law in general relativity:
\begin{equation}
T({\bf r})= \frac{T}{\sqrt{-g_{00}({\bf r})}}\,,
\label{Tolman}
\end{equation}
where the effective metric is
\begin{equation}
-g_{00}({\bf r})=1-\frac{{\bf v}_s^2({\bf r})}{c^2}\,,
\label{EffectiveMetric}
\end{equation}
and $c$ is the speed of sound.
Later, such effective metric experienced by sound waves in liquids (or correspondingly phonons in superfluids)  became known as the acoustic metric,\cite{Unruh1981} and the flow of the liquid with the acoustic horizon as the river model of black holes.\cite{RiverModel}
 
 In general relativity (GR), the flow metric with acoustic horizon corresponds to the Painlev\'e-Gullstrand (PG) metric.  \cite{Painleve,Gullstrand} This suggested that the gravitational field of the black hole with mass $M$ can be considered as the result of the flow of the quantum vacuum with superfluid velocity $v_{\rm s}^2= 2GM/r$ with the event horizon at $v_s^2=c^2$. It is important that the PG metric is continuous across the horizon, which allows to study the thermodynamics of the vacua with horizons. Both for the real black hole and for  its condensed matter analog, the Hawking radiation can be considered as semiclassical quantum tunneling across the horizon, see Ref. \cite{Volovik1999} for the analog black hole and  
Refs. \cite{ParikhWilczek2000,Srinivasan1999}
 for the real black hole. 

The fact that the speed of the "vacuum" exceeds the speed of light beyond the horizon does not mean that the laws of physics are violated. The speed of light remains the speed limit for particles moving relative to the vacuum.
Similarly, a superfluid liquid can move at a speed significantly greater than the speed of sound, but the relative speed of phonons moving relative to the superfluid "vacuum" is the speed of sound.

 \subsection{Tetrad gravity from topological Weyl point}
 \label{WeylPointSec}
 
Acoustic gravity is the part of the so-called analogue gravity, which includes also the other types of condensed matter systems, such as elastic media\cite{Bilby1956,Kroener1960,Dzyaloshinskii1980,KleinertZaanen2004,Vozmediano2010,deJuan2010,Zaanen2010} and topological matter with Weyl fermions. The relativistic Weyl fermions in semimetals were considered by Abrikosov and Beneslavskii\cite{Abrikosov1971,Abrikosov1972,Abrikosov1998}
 (both from Landau Institute). The Weyl materials allow us to simulate the horizon of black and white holes by tilting the Weyl cone.\cite{Volovik2016c,Wilczek2020,Baskaran2025} 

 In the fermionic Weyl and Dirac materials, emergent gravity is formulated in terms of tetrad fields, instead of the metric gravity emerging in bosonic condensed matter systems. 
The tetrad gravity emerges together with all the ingredients of the  relativistic quantum field theories (relativistic spin, chiral Weyl fermions, gauge fields,
$\Gamma$-matrices, etc.).\cite{Volovik2003}
 The spin of Weyl fermions in particle physics and pseudo-spin in Weyl materials forms the hedgehog in momentum space -- the Weyl point in the fermionic spectrum. It represents the Berry phase monopole, which acts as a source or a sink of the Berry curvature.\cite{Volovik1987b} The stability of this hedgehog is supported by topology in momentum space.  It is the topological stability of the Weyl point, which provides the emergence of the relativistic physics at low energy.

 \subsection{Supeprlastic vacuum and translational gauge fields}
  \label{SuperlasticSec}
  
The theory of elasticity in crystals can be considered in terms of elasticity tetrads.\cite{Dzyaloshinskii1980} These tetrads represent the so-called translational gauge fields, which are expressed in terms of a system of deformed crystallographic coordinate planes, surfaces of constant phase,\cite{NissinenVolovik2018b,NissinenVolovik2019,NissinenHeikkila2021,Nissinen2022}   $X^a(x)=2\pi n^a$:
\begin{equation}
 E^{~a}_\mu(x)= \partial_\mu X^a(x)\,.
\label{ElasticityTetrads}
\end{equation}
It is important that such gravitational tetrads have dimensions of inverse length and inverse time, 
$[E^{~a}_i]=[L]^{-1}$ and $[E^{~a}_0]=[t]^{-1}$.
 When these elasticity tetrads are applied to general relativity (the so called superplastic vacuum \cite{KlinkhamerVolovik2019}), one obtains that the Ricci curvature scalar ${\cal R}$ is dimensionless. The same dimensions of tetrads, $[E^{~a}_i]=[L]^{-1}$ and $[E^{~a}_0]=[t]^{-1}$, takes place in the Diakonov theory discussed in Sec. \ref{DIakonovDec}.

 \subsection{Tetrads as bilinear combination of fermions}
 \label{DIakonovDec}
 
Another condensed matter example of effective gravity is provided by the B-phase of superfluid $^3$He, where vielbein emerge as bilinear combinations of the fermionic fields.\cite{Volovik1990} Similar mechanism of the formation of the composite tetrads in the low energy physics has been suggested in the relativistic quantum field theories,\cite{Akama1978,Diakonov2011,VladimirovDiakonov2012,VladimirovDiakonov2014,ObukhovHehl2012,Maiezza2022,Maitiniyazi2025} see Sec. \ref{CompositeTetradSec}. 
The emergent tetrads give rise to the effective metric (the four fermions object), to the interval, and finally to the effective action for the gravitational field. 
As in the case of elasticity tetrads in Eq.(\ref{ElasticityTetrads}), the composite gravitational tetrads also have the unusual dimensions. The consequences of such dimensions for physics are discussed in Sec. \ref{PlanckConstantsSec}

\section{de Sitter state as heat bath}
\label{HeatBathSec}

\subsection{de Sitter state vs moving vacuum}
\label{MovingVacuumSec}

We consider the de Sitter thermodynamics using the Painlev\'e-Gullstrand (PG) coordinates,\cite{Painleve,Gullstrand} where the metric is
 \begin{eqnarray}
ds^2= - c^2dt^2 +   (d{\bf r} - {\bf v}({\bf r})dt)^2 =
\nonumber
\\
= -c^2\left(1 -\frac{{\bf v}^2({\bf r})}{c^2}\right)-2{\bf v}({\bf r})\cdot d{\bf r}\,dt +d{\bf r}^2\,.
\label{PG1}
\end{eqnarray}
Here ${\bf v}({\bf r})$ is the shift velocity, which in condensed matter plays the role of the superfluid velocity ${\bf v}_s$ in Eq.(\ref{EffectiveMetric}) -- the velocity of the "superfluid quantum vacuum".\cite{Volovik2003}
In the de Sitter expansion the velocity of the "vacuum" is ${\bf v}({\bf r})=H{\bf r}$, where $H$ is the Hubble parameter, and the metric is (we use $c=1$):
 \begin{equation}
ds^2= - dt^2 +   (dr - Hr dt)^2+r^2 d\Omega^2 \,.
\label{PG}
\end{equation}
 
The PG metric is stationary, i.e. does not depend on time, and it does not have the unphysical singularity at the cosmological horizon. That is why it is appropriate for consideration of the local thermodynamics both inside and outside the horizon. It also allows us to consider two different phases of the vacuum with broken time reversal symmetry. These are the expanding de Sitter Universe with $H>0$ and the contracting de Sitter Universe with $H<0$. These two degenerate states  transform to each other under the time reversal, $t\rightarrow -t$. In this sense, the Hubble parameter $H$ can be considered as the order parameter of the symmetry breaking phase transition from the symmetric state -- the Minkowski vacuum with $H=0$.

\subsection{Hydrogen atom in de Sitter environment and de Sitter temperature}
\label{AtomInDS}

Let us now show that matter perceives the de Sitter state of the quantum vacuum as the heat bath. For that  let us consider an atom  at the origin, $r = 0$. The atom plays the role of the detector (or the role of the static observer) in this spacetime. The electron bounded to an atom may absorb the energy from the gravitational field of the de Sitter background and escape from the electric potential barrier.  If the ionization potential is much smaller than the electron rest energy but is much larger than the Hubble parameter, $\hbar H\ll \epsilon_0 \ll mc^2$, one can use
the non-relativistic quantum mechanics to estimate the tunneling rate through the barrier. 

Let's consider an electron at the $n$-th level in a hydrogen atom. Under de Sitter gravitational field this electron can escape from the atom with the conservation of energy, which in the classical limit is given by the classical equation: 
\begin{eqnarray}
\frac{p_r^2}{2m} +p_rv(r) = -E_n \,\,,\,\, E_n= \frac{me^4}{2\hbar^2}\frac{1}{n^2}\,.
\label{Classical1}
\end{eqnarray}
 Here $v(r)=Hr$ and $p_r(r)v(r)$ is the Doppler shift, which allows for electron to reach the negative energy when it escapes from the atom.
 The corresponding radial trajectory $p_r(r)$ for escape of electron from the atom is:
\begin{eqnarray}
p_r(r)= -mv(r) + \sqrt{m^2v^2(r) -2m E_n}\,.
\label{ElectronTrajectory}
\end{eqnarray}
The integral of $p_r(r)$ over the classically forbidden region, $0 < r < r_n=\sqrt{2E_n/mH^2}$, gives the  ionization rate 
\begin{eqnarray}
w\sim \exp{ \left(-\frac{2}{\hbar}\,{\rm Im}\, S\right)}=
\nonumber
\\
= \exp{ \left(-\frac{2}{\hbar}\int_0^{r_n} dr \sqrt{2m E_n-m^2H^2r^2}\right)} =
\label{WKB}
\\
=\exp\left(-\frac{\pi E_n}{\hbar H} \right) \equiv exp\left(-\frac{E_n}{T} \right) \,\,,\,\, T=\frac{\hbar H}{\pi}\,.
\label{IonizationRate}
\end{eqnarray}
The ionization rate is equivalent to the rate of ionization in the flat Minkowski spacetime in the presence of the heat bath with temperature $T=\hbar H/\pi$. 
This suggests that the de Sitter state of the quantum vacuum serves as the heat bath for matter.

This heat bath temperature is twice the Gibbons-Hawking temperature  $T_{\rm GH}=H/2\pi$, which is generally considered as the temperature of the cosmological horizon. Since the electron's trajectory is deep inside the horizon, $r_n \ll r_H=c/H$, the ionization process is fundamentally different from the process of Hawking radiation from the cosmological horizon. However, there is a relation between the ionization temperature $T$ and the Gibbons-Hawking temperature of Hawking radiation, $T=2T_{\rm GH}$. This follows from the symmetry of the de Sitter state.\cite{Volovik2024}

The ionization rate can be obtained using a simpler method.\cite{Maxfield2022} Since this process takes place well inside the horizon, one can use the classical (i.e. non-relativistic) gravitational potential $U(r)=-mv^2(r)/2=-mH^2r^2/2$.
Then the bound state decays by quantum tunnelling of electron from the point $r=0$ to the point $r=r_n$, at which the electron level  $-E_n$ matches the de Sitter gravitational potential,  $U(r_n)=-mH^2r_n^2/2=-E_n$. The radial trajectory $p_r(r)$ follows now from the classical equation 
\begin{equation}
\frac{{\bf p}^2}{2m}  - \frac{1}{2}mH^2r^2 = -E_n\,.
\label{GravPotential}
\end{equation}
One obtains
\begin{eqnarray}
p_r(r)= \sqrt{m^2H^2r^2 -2m E_n}\,,
\label{ElectronTrajectory2}
\end{eqnarray}
which again gives Eq.(\ref{IonizationRate}) for the WKB tunneling rate of ionization of atom.

\subsection{Proton decay in de Sitter environment}
\label{ProtonSec}

The same temperature $T=H/\pi$  determines the other processes which are forbidden in Minkowski vacuum, but are allowed in the de Sitter state. This includes for example the decay of proton in the de Sitter environment, which represents the inverse $\beta$-decay, $p^+ \rightarrow n + e^+ +\nu$. The decay rate of proton is:\cite{Volovik2024d}
\begin{eqnarray}
\Gamma(p^+ \rightarrow n + e^+ +\nu) \sim  \exp{\left(-\frac{\pi\Delta M}{H} \right)}
= \exp{\left(-\frac{\Delta M}{T} \right)},\,\,
\label{ProtonDecayRate}
\\
\Delta M = M_n +M_e + M_\nu -M_p\,.\,\,\,
\label{MassDifference}
\end{eqnarray}
Here $M_n$, $M_e$ and $M_p$ are correspondingly masses (rest energies) of neutron, electron and proton, and $M_\nu$ corresponds to the mass eigenstates for neutrino.

This in turn leads to the multiple creation of matter in the de Sitter heat bath.
The created neutron experiences the $\beta$-decay with creation of proton, electron and antineutrino. The 
created proton again experiences the inverse $\beta$-decay, and so on. This leads to the multiple creation of the electron-positron pairs. The creation of matter leads to the decrease of the vacuum energy and thus to the relaxation of the de Sitter state towards the final Minkowski vacuum state. That is why even a single proton in the de Sitter environment triggers the instability of the de Sitter vacuum. This provides the route to the solution of the cosmological constant problem discussed in Section \ref{CosmologicalProblem}. 

\subsection{Local temperature and local entropy}
\label{LocalT}

The behaviour of matter in the de Sitter environment suggests that the de Sitter state can be considered as the heat bath, where the local temperature 
$T=H/\pi$ determines the thermodynamics of the de Sitter state.

From Friedmann equations of general relativity one can express the energy density of the de Sitter vacuum in terms of this local temperature, and then find its free energy and the entropy density.
The energy density, which is the cosmological constant $\Lambda$, is ($\hbar=c=1$):\cite{Volovik2024}
\begin{equation}
 \epsilon_{\rm vac}=\Lambda=\frac{3}{8\pi G}H^2=\frac{3\pi}{8G}T^2\,.
\label{dSEnergyDensity}
\end{equation}
This determines the free energy density $F$ of the de Sitter state. From equation $F- T dF/dT=\epsilon_{\rm vac}$ one obtains $F(T)=-\epsilon_{\rm vac}(T)$, and thus the entropy density $s_{\rm dS}$ is:
\begin{equation}
s_{\rm dS}= - \frac{\partial F}{\partial T} =\frac{3\pi}{4G}T=\frac{3}{4G}H\,.
\label{dSEntropyDensity}
\end{equation}

This is an example of the entropy density of spacetime suggested by Padmanabhan,\cite{Padmanabhan2010} see also review papers on the gravitational entropy.\cite{Guha2023,Ong2022}

\subsection{de Sitter thermodynamics and holography}
\label{holography}

 Now let us consider the possible connection of this local thermodynamics of de Sitter state with the global thermodynamics of black holes, where the total entropy of the black hole is determined by the area of its horizon. For that let us find the entropy $S_H$ of the part of the de Sitter space, which is surrounded by the cosmological horizon -- the total entropy of the Hubble volume $V_H=(4\pi/3) \,r_H^3$. Multiplying the entropy density in Eq.(\ref{dSEntropyDensity}) by the Hubble volume one obtains
  \begin{equation}
S_H=V_H s_{\rm dS}= \frac{A}{4G}\,,
\label{dSEntropyHubble}
\end{equation}
where $A=4\pi r_H^2$ is the horizon area. This bulk entropy exactly coincides with the entropy of the cosmological horizon suggested by Gibbons and Hawking. However, this global entropy comes from the local entropy of the de Sitter state, rather than from the horizon degrees of freedom. This demonstrates the specific features of the de Sitter horizon in relation to holography discussed e.g. in Refs.\cite{Padmanabhan2010,Susskind2021,Bobev2023,Zenoni2024}.
Equation (\ref{dSEntropyHubble}) confirms the holographic bulk-horizon correspondence in the de Sitter state, which in turn confirms the local thermodynamics of the de Sitter state with the double Hawking temperature, $T=H/\pi=2T_{\rm GH}$.

\subsection{Double Hawking temperature measured by freely-falling detector}
\label{FreeFalling}

The doubling of the Hawking temperature was suggested for black holes by 't Hooft.\cite{Hooft2022,Hooft2023,Hooft2024}  In the 't Hooft scenario, the doubling of temperature is accompanied by the reduction of the total entropy of the black hole by a factor 2. In the de Sitter case, the temperature is also twice the usually accepted value, but the entropy in the equation (\ref{dSEntropyHubble}) has the traditional value expected from the holographic principle.

Another example of the doubling of the Hawking temperature of black hole, which was discussed in Ref. \cite{Carroll2025}, is rather close to the de Sitter scenario. It is now the effective temperature measured by a freely-falling observer crossing the black hole horizon, which is twice the Hawking temperature. The temperature of Hawking radiation from the black hole and the effective temperature measured by a freely-falling detector at the black hole horizon are obtained in the same quantum tunneling approach, where the radiation rate is:
\begin{eqnarray}
w\sim \exp{ \left(-\frac{2}{\hbar}\,{\rm Im}\, S\right)}\,,
\nonumber
\\
{\rm Im}\, S ={\rm Im} \int dr \,\frac{E}{1+v(r)}= \frac{\pi E }{|dv/dr|_{r=r_H}} \int_{r_1} ^{r_2}dr\,\delta (r-r_H)\,.
\label{WKBHawking}
\end{eqnarray}
Here $r_H$ is the position of the event horizon.

In the case of Hawking radiation, we have $r_1 < r_H< r_2$, and the integral $\int dr\,\delta (r-r_H)=1$. As a result, we obtain the conventional Hawking temperature, $T_H=|dv/dr|_{r=r_H}/2\pi$. In the case of the detector crossing the horizon, the radiation is caused by the detector itself. The detector plays the role of an external object that produces radiation in the same way as an external atom provides radiation of an electron in a de Sitter environment. In this case, $r_1$ is determined by the position of the detector, $r_1=r_H <r_2$, and the integral is
$\int dr\,\delta (r-r_H)=1/2$. As a result, the local temperature measured by the detector is twice as large, $T=|dv/dr|_{r=r_H}/\pi =2T_H$.

\section{Two-component thermodynamics of the de Sitter state}
\label{dSthermodynamicsSec}

\subsection{Modified Gibbs-Duhem relation, dark energy and dark matter}
\label{GDrelation}

The quadratic dependence of vacuum energy on temperature in Eq.(\ref{dSEnergyDensity}) is important for consideration of the thermodynamic Gibbs-Duhem relation for quantum vacuum. It leads to the reformulation of the vacuum pressure.
In the conventional approach, the vacuum pressure $P_{\rm vac}$ obeys the equation of state $P=w\epsilon$ with $w=-1$, i.e.  $P_{\rm vac}=-\epsilon_{\rm vac}$ (let us recall that for non-relativistic matter $w=0$ and for radiation $w=1/3$).  In the de Sitter state the vacuum pressure  is negative, $P_{\rm vac}=-\epsilon_{\rm vac}<0$. 
This pressure $P_{\rm vac}$ does not satisfy the thermodynamic Gibbs-Duhem relation, $Ts_{\rm dS}=  \epsilon_{\rm vac}+ P_{\rm vac}$, because the right hand side of this equation is zero.
The reason for that is that in this equation we did not take into account the gravitational degrees of freedom. 

It was shown, that gravity contributes to thermodynamics with the pair of the  thermodynamically conjugate variables:  the gravitational coupling $K=\frac{1}{16\pi G}$ and the Riemann curvature  ${\cal R}$, see Refs.\cite{KlinkhamerVolovik2008c,Volovik2022G,Volovik2020}. The gravitational thermodynamic variables allow us to write the modified  Gibbs-Duhem relation:
 \begin{equation}
Ts_{\rm dS}=  \epsilon_{\rm vac}+ P_{\rm vac} -K{\cal R}\,.
\label{GibbsDuhem}
\end{equation}
This equation allows us to introduce a pressure that characterizes the gravitational degrees of freedom:
\begin{equation}
P_{DM}= P_{\rm vac} -K{\cal R} \,.
\label{EffectiveP}
\end{equation}
Then the Gibbs-Duhem relation becomes:
\begin{equation}
Ts_{\rm dS}=  \epsilon_{\rm vac}+ P_{DM}\,.
\label{EffectiveGibbs}
\end{equation}
We call $P_{DM}$ the pressure of the dark matter component, although it has nothing to do with real physical dark matter.

Assuming that the entropy of de Sitter comes from the gravitational degrees of freedom, i.e. $s_{\rm dS}=s_{\rm DM}$, one obtains from Eq.(\ref{EffectiveGibbs}) that the energy density of this `dark matter' equals the energy density of the vacuum, $\epsilon_{\rm DM}=\epsilon_{\rm DE}$.
That is why our dark matter satisfies the equation of state $P=w\epsilon$ with $w=1$. This corresponds to stiff matter introduced by Zel'dovich,\cite{Zeldovich1962} where the speed of sound is equal to the speed of light, $c_s^2=c^2 dP/d\epsilon_{\rm vac}=c^2$.

\subsection{Two dark components of de Sitter and  Zel'dovich stiff matter}
\label{StiffMatterSec}

So, the thermodynamics of the de Sitter expansion can be described in terms of two components of the de Sitter state: the vacuum component (dark energy, DE) and the stiff matter (dark matter, DM).
These components have the following forms:
\begin{equation}
w=-1\,\,,\,\,  P_{\rm DE}(H)= - \epsilon_{\rm DE}(H)=-\frac{3}{8\pi G}H^2\,,
\label{DarkEnergy}
\end{equation}
and
\begin{equation}
w=1\,\,,\,\, P_{\rm DM}(T)=  \epsilon_{\rm DM}(T)=\frac{3\pi}{8G}T^2\,,
\label{DarkMatter}
\end{equation}

Then from equations (\ref{GibbsDuhem})-(\ref{EffectiveGibbs}) it follows that in equilibrium, i.e. at $T=H/\pi$, the total (thermodynamic) pressure is zero, as it should be in the absence of the external environment when the partial pressure of dark matter compensates the partial pressure of quantum vacuum:
\begin{equation}
P=P_{\rm DE}(H=\pi T)+P_{\rm DM}(H=\pi T)=  0\,.
\label{PressureVacuum}
\end{equation}
In this equilibrium state, the energy density of dark matter is equal to the dark energy density:
\begin{equation}
\epsilon_{\rm DM}(H=\pi T)=\epsilon_{\rm DE}(H=\pi T)\,.
\label{Equilibrium}
\end{equation}
The modified Giggs-Duhem relation in Eq.(\ref{EffectiveGibbs}) represents the Gibbs-Duhem relation for dark matter:
\begin{equation}
Ts_{\rm DM}=  \epsilon_{\rm DM}+ P_{DM} \,\,,\,\, s_{\rm DM}\equiv s_{\rm dS}\,.
\label{EffectiveGibbs2}
\end{equation}
In other words, the thermodynamics of the de Sitter state is fully represented by the thermodynamics of the gravitational dark matter, which plays the role of the normal component in this two-fluid thermodynamics.

So, we have the following set of equations for the equilibrium de Sitter state, which is expressed in terms of the two dark components:
\begin{eqnarray}
P_{\rm DE}=-\epsilon_{\rm DE}\,,
\label{Equilibrium1}
\\
P_{\rm DM}=\epsilon_{\rm DM}\,,
\label{Equilibrium2}
\\
P=P_{\rm DE}+P_{\rm DM}=  0 \,,
\label{Equilibrium3}
\\
Ts_{\rm DM}=  \epsilon_{\rm DM}+ P_{DM}\,.
\label{Equilibrium4}
\end{eqnarray}
Since the dark energy component represents the vacuum, it acquires the nonzero value only for compensation of the dark matter pressure  in Eq.(\ref{Equilibrium3}) in equilibrium. In the absence of environment, the external pressure is zero, $P=0$. Because of that the energies of the two components in equilibrium have equal values in Eq. (\ref{Equilibrium}).

The similar situation takes place for the closed static Universe with positive curvature introduced by Einstein. In this case there are three contributions to the pressure: from the ordinary matter, from the vacuum energy and from the gravitational degrees of freedom (i.e. from the spatial curvature $\cal R$), see Sec. 29.4.5 in the book \cite{Volovik2003} and Ref. \cite{Volovik2024e}. Since the static Universe has no environment, in the full equilibrium the total pressure is zero,  $P=P_{\rm vac} + P_{\rm matter}+P_{\rm gravity}=0$. The equilibrium conditions provide the connection between matter, dark energy and curvature of the Einstein Universe.

\subsection{Zel'dovich stiff matter and Landau Fermi liquid}
\label{LandauFermiSec}

The equation (\ref{dSEntropyDensity}) demonstrates that the thermal properties of the de Sitter state are similar to that of the non-relativistic Fermi liquid, such as liquid $^3$He. The entropy density in the Fermi liquid is also linear in temperature:
\begin{equation}
s_{\rm FL}=\frac{p_F^2}{3v_F}T\,.
\label{FLEntropyDensity}
\end{equation}
The Fermi velocity $v_F$ and Fermi momentum $p_F$ of this cosmological analog of the Fermi liquid are on the order of the speed of light and the inverse Planck length correspondingly, $v_F \sim c$ and 
$p_F\sim 1/l_{\rm P} \equiv M_{\rm P}/c$, where $M_{\rm P}$ is the reduced Planck mass, $M_{\rm P}^2=1/(8\pi G)$. 

The Sommerfeld law in Fermi liquid states that the entropy per one atom of the Fermi liquid is 
\begin{equation}
S=\frac{s_{\rm FL}}{n_{\rm FL}}\sim \frac{T}{E_F}\,,
\label{FLEntropyDensity}
\end{equation}
 where $n_{\rm FL} \sim p_F^3$ is the density of atoms in the Fermi liquid and $E_F$ is Fermi energy. 
 
 Can it be related to quantum vacuum?
  We do not know what are  the "atoms of the vacuum", but from Eq.(\ref{dSEntropyDensity}) it follows that the entropy density of the vacuum  
$s_{\rm vac} \sim T/l_{\rm P}^2 \sim (T/M_{\rm P})/l_{\rm P}^3$.
This suggests that the density of the "atoms of the vacuum" is $n_{\rm P} \sim 1/l_{\rm P}^3$ and entropy per  "atom of the vacuum" is:
\begin{equation}
S= \frac{s_{\rm vac}}{n_{\rm P}} \sim s_{\rm vac} l_{\rm P}^3 \sim \frac{T}{M_{\rm P}}  \,.
\label{Sommerfeld}
\end{equation}
  Eq.(\ref{Sommerfeld}) is the full analog of the Sommerfeld law for Fermi liquids. In the Fermi liquids this corresponds to the density of states at the Fermi level $N_F\sim mp_F$. This analogy suggests that the corresponding density of states in the quantum vacuum is $N_{\rm P} \sim M_{\rm P}^2$.  The huge density of states in the quantum vacuum leads to a very large entropy of the de Sitter state even at a very small temperature of the vacuum.

\section{Two-component dynamics of the de Sitter state}
\label{dSdynamicsSec}

\subsection{Phenomenological theory of two-component dynamics}
 \label{PhenomenologySec}

The two-fluid thermodynamics of the de Sitter state discussed in Sec. \ref{dSthermodynamicsSec} is valid in thermal equilibrium, when $T=H/\pi$. 
However, the equilibrium state is impossible since the gravitational dark matter represents the heat bath for the ordinary matter,
which leads to the thermal nucleation of matter and thus to decay of the vacuum energy. Here we will not go into the details of this decay process, which depends on many different factors. 
We shall use the phenomenological approach suggested in Ref.\cite{Volovik2024} and apply to the Universe with the cosmological stiff matter.  We assume that the Einstein equations are modified in such a way that  the gravitational dark matter behaves as the real  Zel'dovich stiff matter also in dynamics.
Then the dark matter decays with time due to expansion, but it also gets energy from the energy exchange with the dark energy. On the other hand this energy exchange leads to decrease of the vacuum energy density, and the Universe finally approaches the equilibrium Minkowski vacuum state. 

So, let us consider how the de Sitter state relaxes if one takes into account that the temperature of dark matter changes both due to the expansion and due to the energy exchange with the dark energy. The only assumption in this phenomenological approach is that  due to the energy exchange, both dark components try to reach the common temperature in the process of equilibration.

\subsection{Dynamics without energy exchange}
 \label{NoEnergyExchangeSec}

If there is no energy exchange between the two components, the real stiff matter with equation of state $w=1$ relaxes according to the Einstein equations:
\begin{eqnarray}
\partial_t \epsilon_{\rm DM}= - 3H(w+1) \epsilon_{\rm DM}=- 6H \epsilon_{\rm DM}\,,
\label{Stiff}
\end{eqnarray}
while the equation for the vacuum energy density with $w=-1$ is time independent:
\begin{eqnarray}
\partial_t \epsilon_{\rm DE}=- 3H(w+1) \epsilon_{\rm DE}=0\,.
\label{deSitter}
\end{eqnarray}
Of course, for the original gravitational stiff matter one has $\partial_t \epsilon_{\rm DM}=0$ instead of Eq.(\ref{Stiff}). But here we consider the toy phenomenological model, in which the real Zeldovich matter has exactly the same energy density as the gravitational Zeldovich matter in Eq.(\ref{DarkMatter}).

\subsection{Dynamics with the energy exchange}
 \label{EnergyExchangeSec}

If there is the energy exchange between the two components, it is natural to assume, that these dark components tend to approach the common equilibrium state in Eq.(\ref{Equilibrium}). This suggests that due to the energy exchange in the process of thermalization, the temperature of the dark matter tends to approach the temperature $T=H/\pi$ of the de Sitter thermal bath, and thus the energy density of dark matter tends to approach the dark energy density, $\epsilon_{\rm DM}(T)\rightarrow\epsilon_{\rm DE}(H)$.
For that the equation (\ref{Stiff}) must be modified in the following way:
\begin{eqnarray}
\partial_t \epsilon_{\rm DM}(T)= - 6H (\epsilon_{\rm DM}(T)- \epsilon_{\rm DE}(H))\,.
\label{StiffNonConservation}
\end{eqnarray}
This is the phenomenological equation, which describes the process of the thermalization by the energy exchange between the two dark components.

Now, the gain of $6H\epsilon_{\rm DE}(H)$ in the energy of dark matter in Eq.(\ref{StiffNonConservation}) must be compensated by the loss of dark energy. This modifies the Eq. (\ref{deSitter}), leading to a relaxation of the vacuum energy:
\begin{eqnarray}
\partial_t \epsilon_{\rm DE}(H)= -6H \epsilon_{\rm DE}(H) \,.
\label{VacuumNonConservationS}
\end{eqnarray}

The equation (\ref{VacuumNonConservationS}) gives the following asymptotic dependence of dark energy at large times:
\begin{eqnarray}
\epsilon_{\rm DE}(t) =\frac{1}{6\pi Gt^2} \,.
\label{DecayLaw}
\end{eqnarray}
The same behaviour of the vacuum energy density was obtained using the Hawking 4-form field,\cite{Hawking1984} see Refs.\cite{KlinkhamerVolovik2008c,Volovik2013}. In this case the role of dark matter is played by the oscillations of the vacuum energy during its decay.\cite{KlinkhamerVolovik2017} 

From Eq.(\ref{StiffNonConservation}) one obtains the time dependence of dark matter:
\begin{eqnarray}
\epsilon_{\rm DM}(t) =\frac{ \ln(t/t_0)}{3\pi Gt^2}\,.
\label{DecayLawMatter}
\end{eqnarray}
Both dark components are of the same order of magnitude and have reasonable values at present:
\begin{equation}
 \epsilon_{\rm DE}(t_{\rm present})\sim \epsilon_{\rm DM}(t_{\rm present})\sim H^2M^2_{\rm P} \sim 10^{-120} M^4_{\rm P}
\,.
\label{PresentTime}
\end{equation}
Both components finally approach the fully equilibrium state with zero temperature -- the Minkowski vacuum.  That is why this toy phenomenological model shows the route to the solution of the cosmological constant problems including the coincidence problem (see also Section \ref{FineTuningSec}). The key point here is the instability of the de Sitter state in the presence of an external object (for example, a proton).

 \section{Black hole thermodynamics from macroscopic quantum tunneling}
  \label{MacroscopicTunnelSec}

 \subsection{Vortex nucleation as macroscopic quantum tunneling}
  \label{VortexNucllSec}
  
The process of quantum tunneling of macroscopic objects is well known in condensed matter physics, see the original papers from the seventies. \cite{LifshitzKagan1972,IordanskiiFinkelshtein1972,Volovik1972,IordanskiiFinkelshtein1973,IordanskiiRashba1978} Most of the authors of these papers were from the Landau Institute. Macroscopic tunneling can be studied using collective variables that describe the collective dynamics of a macroscopic object. This approach allows one to estimate the semiclassical tunneling exponent without considering the details of the object's structure at the microscopic (atomic) level. That is why it is important in black hole physics, where the structure of the quantum vacuum is still unknown.

The processes of macroscopic quantum tunneling include in particular the process of quantum nucleation of vortices in superfluids.\cite{Volovik1972} 
In this process, the role of the canonically conjugate collective variables is played by the $z$ and $r$ coordinates of the vortex ring. This provides the volume law for the vortex instanton: the action contains the topological term $S_{\rm top}=2\pi \hbar n V=2\pi \hbar N$, where $n$ is the particle density; $V$ is the volume inside the surface swept by vortex line between its nucleation and annihilation; and $N$ is the number of atoms inside this volume, see Sec. 26.4.3 in the book \cite{Volovik2003} and also the paper by Rasetti and Regge.\cite{Regge1975}

 For the other linear topological defects in condensed matter the area law
is applied instead of the volume law, i.e. the action is proportional to the area of the surface swept by the defect line, as in the case of the Polyakov string action.\cite{Polyakov1981} The application of the macroscopic quantum tunneling to Abrikosov vortices in superconductors can be found in the review paper,\cite{Blatter1994} where most authors were from the Landau Institute.

The rate of the emission of the vortex loop in the superfluid helium moving with respect to the walls is
\begin{equation}
w \propto \exp{\left(-2\pi N\right)} 
\,,
\label{VortexNucleation}
\end{equation}
where $N$ the number of "atoms of the vacuum" (atoms of $^4$He) participating in the vortex instanton. It depends on the "velocity of the vacuum", $N\propto 1/v_s^3$. This is similar to the quantization of the black hole 
entropy in terms of micro-states of Planck size, where the number of "black hole atoms" is determined by the ratio between mass $M$ of black hole and the Planck mass $M_{\rm P}$. In the black hole case the tunneling rate is related to the change of the entropy after tunneling, which gives the rate of emission
in terms of the number of microstates similar to that in Eq. (\ref{VortexNucleation}). This will be discussed in Sec. \ref{BHensemble}.

 \subsection{Hawking radiation as quantum tunneling}
\label{HawkingRadiation}

Let us start with the quantum tunneling of particles from the black hole, which describes the Hawking radiation.\cite{ParikhWilczek2000,Srinivasan1999,Volovik1999}
The rate of emission of the particle with energy $\omega$ from the black hole with mass $M$ has the following exponential law:
\begin{equation}
w(\omega, M)\propto   \exp{\left(-8\pi GM\omega\right)}\equiv \exp{\left(-\frac{\omega}{T_{\rm H}}\right)} 
\,.
\label{tunneling}
\end{equation}
Here $T_{\rm H}$ is the temperature of the Hawking radiation:
\begin{equation}
T_{\rm H}=\frac{1}{8\pi GM}
\,.
\label{HawkingT}
\end{equation}
We use $\hbar=c=1$ and the condition $T_{\rm H}\ll \omega \ll M$.
Thus the quantum tunneling process demonstrates that  the rate of emission is determined by the Hawking temperature $T_{\rm H}$. This is important for the construction of the black hole thermodynamics.

The further important step was made by Parikh and Wilczek,\cite{ParikhWilczek2000} who obtained the correction to the Hawking radiation. This correction is caused by the back reaction --  the reduction of the black hole mass after emission:
 \begin{equation}
w(\omega, M-\omega)\propto \exp{\left(-8\pi G\omega\left(M-\frac{\omega}{2}\right)\right)}
\,.
\label{tunnelingMomega}
\end{equation}

This back reaction can be related to the decrease of the Bekenstein-Hawking entropy $S_{\rm BH}(M)=4\pi GM^2$ of the black hole after emission.\cite{Kraus1997} The reason is that the Hawking radiation can be considered as the rare effect caused by thermodynamic fluctuations. According to Landau and Lifshitz,\cite{Landau_Lifshitz} the rare fluctuations lead to the decrease of entropy, and thus
the rate of Hawking radiation in Eq.(\ref{tunnelingMomega}) can be described in terms of the decrease of the  black hole entropy after emission of a particle:
\begin{equation}
w(\omega, M-\omega)\propto \exp{\left[S_{\rm BH}(M-\omega)-S_{\rm BH}(M)\right]}
\,.
\label{EntropyDifference}
\end{equation}
This shows that the quantum tunneling process may serve as a source of both the Hawking temperature and the Bekenstein-Hawking entropy. This is supported by consideration of macroscopic quantum tunneling discussed in Sec. \ref{TunnelingEmission}. It describes the process in which a large black hole emits smaller black holes.\cite{Volovik2022G}  

Moreover, the macroscopic quantum tunneling from the black hole to the white hole with the same mass demonstrates the curious thermodynamic properties of the white hole.\cite{Volovik2025P}  In particular, the entropy of the white hole is negative, $S_{\rm WH}(M)=-S_{\rm BH}(M)$. This is a consequence of a special type of the white hole event horizon that also occurs in a collapsing de Sitter universe with $H<0$. It is unclear whether negative entropy can arise in systems without an event horizon and in condensed matter analogues.

\subsection{Emission of black holes as macroscopic quantum tunneling}
\label{TunnelingEmission}

The emitted black hole can be thought of as a type of particle, and the emission process contains a similar element of back reaction.\cite{Volovik2022G} However, such "particle" emitted by the black hole has a non-zero entropy. As a result, unlike the emission of a point particle, the emission rate of a small black hole increases by the entropy of the emitted black hole.\cite{HawkingHorowitz1995} 

All this suggests that the general process of the splitting of black holes into several parts can be expressed in terms of the entropies of the black holes participating in this process. In particular, the rate at which a black hole splits into two smaller black holes in the process of macroscopic quantum tunneling obeys the following rule:
 \begin{eqnarray}
w(M\rightarrow M_1+M_2)\propto 
\nonumber
\\
\propto\exp{\left[S_{\rm BH}(M_1)+S_{\rm BH}(M_2)-S_{\rm BH}(M_1+M_2)\right]}
\,.
\label{HoleEmession2}
\end{eqnarray}
This equation allows us to obtain the entropy of the black hole as a function of its mass, 
$S_{\rm BH}(M)$. For that let us consider the emission of a small black hole with mass $m$ by a large black hole with mass $M \gg m$.
The rate of emission obeys Eq.(\ref{tunneling}) with $\omega=m$. On the other hand, it obeys Eq.(\ref{HoleEmession2})
with $M_1=M-m$ and $M_2=m\ll M$. Expanding this equation in small $m$ and comparing this equation with Eq.(\ref{tunneling}) one obtains the function $S_{\rm BH}(M)$:
\begin{equation}
\frac{dS_{\rm BH}}{dM}=8\pi GM \,\, \rightarrow \,\, S_{\rm BH}(M)=4\pi GM^2\,.
\label{EntropyExpansion}
\end{equation}

The macroscopic quantum tunneling approach is actually another way to derive the Bekenstein-Hawking entropy of a black hole. Application of the macroscopic quantum tunneling to the  cosmological decay of a false quantum vacuum can be found in Refs.\cite{HawkingMoss1982,Saito2025}.

\subsection{Non-extensive entropy of black hole}
\label{NonextensiveSec}

The black hole entropy is non-extensive, since $S_{\rm BH}(M_1 +M_2) > S_{\rm BH}(M_1)+ S_{\rm BH}(M_2)$. The source of the non-extensive entropy of black holes is a special type of configuration space of the black hole ensemble, which follows from the black hole transformations discussed in Sec. \ref{TunnelingEmission}.

The entropy of the conventional thermodynamic systems is extensive. The entropy there is proportional to the volume of the system, and thus the splitting of the system with volume $V$ in two parts with volumes $V_1+V_2=V$ does not change the total entropy of the system,  $S(V_1+V_2)=S(V_1) +S(V_2)$, or $S(A,B)=S(A) +S(B)$. 

The black hole entropy does not satisfy the additivity condition. Instead one has the following non-additive composition rule for the black hole entropies:
\begin{equation}
S_{\rm BH}(M_1 +M_2)= \left( \sqrt{S_{\rm BH}(M_1)} + \sqrt{S_{\rm BH}(M_2)}\right)^2 \,.
\label{TwoBlackHoles}
\end{equation}

It is precisely because of the quantum processes that the ensemble of black holes has a special type of configuration space, where the entropy is not extensive. This non-extensivity is caused by quantum fluctuations, which determine the rare processes of macroscopic quantum tunneling between the black hole states. This is the analog of intermittency in the chaotic systems.\cite{Robledo2022}
Such processes require the generalization of the statistics with the corresponding non-extensive entropy. The equation (\ref{TwoBlackHoles}) fully determines the type of the statistics: it is described by the Tsallis-Cirto entropy with 
$\delta=2$,\cite{Volovik2025} see Sec. \ref{BHensemble}.

If a black hole is a mixed state and its entropy is the von Neumann entropy, then the quantum tunneling process of breaking the black hole into smaller black holes transforms the mixed state into a less mixed state - a state with lower entropy. 
This is consistent with Weinberg's view\cite{Weinberg2014} that it is the density matrix introduced by Landau and von Neumann that describes physical reality, not the wave function introduced by Schr\"odinger. The wave function approach is as convenient as the complex order parameter (the condensate wave function) introduced by Ginzburg and Landau to describe superfluidity  and superconductivity. But the real physical quantities are their bilinear combinations - the density matrix and the correlation function, which are invariant with respect to phase shift (for additional symmetries of the density matrix compared to the wave function, see Ref. \cite{Weinberg2014}). 

An important example is a two-dimensional superfluid liquid, where the condensate wave function is absent. The phase transition to the superfluid state is determined by the behavior of the correlator.\cite{Kane1967,Berezinskii1972}

\subsection{de Sitter entropy is extensive}
\label{dSextensive}

The non-extensivity of the black hole entropy and hence the application of the generalized statistics is not surprising given the presence of long-range forces.\cite{Odintsov2025}
The generalized statistics is applicable to the black hole entropy as the entropy of the finite closed system. It cannot be applied to such open systems as the de Sitter state, where the entropy is extensive.\cite{Volovik2024,Volovik2024d}
The entropy of any volume $V$ in the de Sitter state, regardless of whether it is smaller or larger than the Hubble volume $V_H$, is proportional to volume:
\begin{equation}
S_V= s_{\rm dS} V = \frac{3\pi}{4G} TV\,.
\label{VolumeEnrropy}
\end{equation}

The extensive entropy in Eq.(\ref{VolumeEnrropy}) determines thermal fluctuations of the thermodynamic variables. 
For example, according to Landau and Lifshitz,\cite{Landau_Lifshitz} temperature fluctuations in a volume $V$ are inversely proportional to the volume. In the de Sitter state one has: 
\begin{equation}
\frac{<(\Delta T)^2>}{T^2} \sim \frac{1}{S_V}=\frac{1}{S_H} \, \frac{V_H}{V}\,.
\label{TFluctuations}
\end{equation}
Here as before $V_H$ and $S_H$ are correspondingly the Hubble volume and the entropy of the Hubble volume.
The extensivity of the de Sitter entropy does not contradict to the holographic bulk-surface correspondence in Eq.(\ref{dSEntropyHubble}), which is applied only to the entropy of the Hubble volume $V_H$. 

\subsection{Entropy of black hole singularity}
\label{SingularitySec}

De Sitter thermodynamics obeys a modified Gibbs-Duhem relation in the equation (\ref{GibbsDuhem}), which includes the gravitational degrees of freedom: the gravitational coupling $K$ (the gravitational analog of the chemical potential) and the scalar curvature ${\cal R}$. Let us assume that a similar relation applies to the local thermodynamics of black holes. In case of the black hole this relation has the following form:
 \begin{equation}
Ts({\bf r})=  \epsilon({\bf r}) -K{\cal R}({\bf r})\,.
\label{GibbsDuhemBH}
\end{equation}
Here the temperature $T$ and the "gravitational chemical potential"  $K$ are constants, while the scalar curvature ${\cal R}({\bf r})$, energy density $\epsilon({\bf r})$ and entropy density $s({\bf r})$ are space dependent. One has:
 \begin{equation}
\epsilon({\bf r}) = M \delta({\bf r}) \,\,,\,\,  {\cal R}({\bf r})=8\pi G M\delta({\bf r})\,.
\label{Singular}
\end{equation}
Then with $T= 1/(8\pi GM)$ and $K=1/(16\pi G)$, the modified Gibbs-Duhem relation (\ref{GibbsDuhemBH}) gives the entropy density, which agrees with the holographic bulk-horizon correspondence:
\begin{eqnarray}
s({\bf r}) = \frac{M}{2T} \delta({\bf r})= \frac{A}{4G} \delta({\bf r})  \,,
\label{SingularEntropy}
\\
S_{\rm BH}=\int d^3r \,s({\bf r}) =\frac{A}{4G} \,.
\label{SingularEntropy2}
\end{eqnarray}

The same singular entropy density in Eq.(\ref{SingularEntropy}) is obtained for the charged Reissner-Nordstr\"om (RN) black hole.
According to Ref.\cite{Volovik2022G} the entropy and the Bekenstein-Hawking temperature of the RN black hole are the same as for the electrically neutral black hole with the same mass. The extra term is to be added to the right hand side of Eq.(\ref{GibbsDuhemBH}), $-\Phi q({\bf r})$, which comes from charge density  
$q({\bf r})=Q\delta({\bf r})$. However this term is zero, because the corresponding electric analogue of the chemical potential (the electrostatic potential)  is zero, $\Phi=0$.

So, if our assumption is correct, then it follows from Eqs. (\ref{SingularEntropy}) and (\ref{SingularEntropy2}) that all the entropy of the black hole, $S_{\rm BH}=A/4G$, is concentrated in the black hole singularity. This is quite natural from the condensed matter point of view: a singularity is a compact physical object with an enormous energy density. The condensed matter analogue of such object is represented by the core of a vortex in a superfluid liquid. The vortex core contains the singularity in the superfluid velocity, 
$\nabla\times {\bf v}_s=\kappa \hat{\bf z}\delta({\bf r}_\perp)$, where $\kappa$ is the quantum of circulation of the superfluid velocity. In the chiral Weyl superfluids, the vortex core has a large density of states, which is caused by the flat band in the spectrum of electrons confined in the singularity.\cite{Volovik2011} 

It would be interesting to apply this approach to the Kerr black hole, and to the super-critical Reissner-Nordstr\"om black hole with naked singularity.\cite{Tumulka2025,Volovik2025P} 

 \section{Unruh effect and Schwinger pair production}
 \label{UnruhSchwingerSec}

\subsection{Schwinger vs Unruh}
 \label{UnruhSchwingerSubSec}

In Section \ref{HeatBathSec} we have seen that the de Sitter state is characterized by the local temperature, which is twice the Gibbons-Hawking temperature.\cite{Volovik2009m,Volovik2024} 
This adds more questions to the many discussions about the factor 2 problem in black hole thermodynamics and in cosmology.\cite{Akhmedov2006,Akhmedov2007,Akhmedov2008,Akhmedova2008,Akhmedova2009,Gill2010}
  The Schwinger pair creation also carries some features of thermal radiation, so one can expect the occurrence of a factor 2 problem in this phenomenon as well.

 The rate of the Schwinger pair creation\cite{Schwinger1951,Schwinger1953} of particles with mass $M$ and charges $\pm q$ in electric field ${\cal E}$ per unit volume per unit time is given by:
\begin{equation}
\Gamma^{\rm Schw}(M) = \frac{dW^{\rm Schw}}{dt} = 
 \frac{q^2{\cal E}^2}{(2\pi)^3}  \exp{\left(- \frac{\pi  M^2}{q\cal E} \right)} \,.
\label{SchwingePairCreation}
\end{equation}
Since $a=q{\cal E}/M$ corresponds to the acceleration of a charged particle, one obtains the analogue of the Unruh effect:\cite{Unruh1976} 
\begin{equation} 
\Gamma^{\rm Schw}(M) \propto \exp{\left(- \frac{\pi  M}{a} \right)} \equiv\exp{\left(- \frac{M}{T} \right)}  =\exp{\left(- \frac{M}{2T_{\rm U}} \right)} .
\label{SchwingerAcceleration}
\end{equation}
The Schwinger pair creation does look as the Unruh effect, although the corresponding effective temperature is twice the Unruh temperature, 
$T=a/\pi \equiv \bar T_{\rm U}=2T_{\rm U}$. There were many attempts to connect these two phenomena, see Refs.\cite{Parentani1997,KharzeevTuchin2005,Kempf2021,Zabrodin2022} and references therein. 

Of course, the similarity between the equations for probabilities suggests that there is some analogy between the Schwinger and Unruh effects. However, the extra factor 2 comes as the serious problem. 
At the Landau Institute, Starobinsky became interested in solving this problem.
Here we consider this problem using the experience with the doubling of the Gibbons-Hawking temperature in case of the cosmological horizon in Section \ref{HeatBathSec}.

If one tries to make the direct analogy between the Schwinger and Unruh processes, this is already problematic. The original state is the vacuum in the constant electric field. Being the vacuum it does not provide any physical acceleration. Acceleration in electric field appears only in the presence of a charged particle. That is why one can try to find the situation, when the two effects are physically connected. The connection may arise if we split the pair creation in several steps. In the first step the pair of particles is created by Schwinger mechanism and then in the further steps the Unruh process enters, which is caused by the acceleration of the created charged particles by electric field. These quantum processes should take place in unison, i.e. coherently. This means that in the quantum tunneling picture,\cite{Volovik1999,ParikhWilczek2000} the corresponding exponents are added.

  \subsection{Double Unruh temperature from Schwinger+Unruh}
 \label{SchwingerDoubleSec}

To connect the Schwinger pair production with the Unruh effect, let us consider the following combined process in which the particles with charges $\pm q$ and masses $M+m$ are created. At first step the particles with charges $\pm q$ and masses $M$ are created, at the second step these charged particles are accelerated by electric field. These particles play the role of the detectors which energy levels are excited from the levels with energy $M$ to the levels with energy $M+m$ due to the Unruh effect. Since these processes occur in unison, the probability of the combined process $\Gamma^{\rm Schw}(M+m)$ is equal to the product of partial probabilities:
 \begin{eqnarray}
\Gamma^{\rm Schw}(M+m) =\Gamma^{\rm Schw}(M)  \, \Gamma_+^{\rm Unruh}(m) \Gamma_-^{\rm Unruh}(m)\,.
 \label{SchwingerUnruhDouble}
 \end{eqnarray}
 Here $\Gamma_+^{\rm Unruh}(m)$ and $\Gamma_-^{\rm Unruh}(m)$ are the excitations rates of the corresponding detectors. 

On the other hand, according to Eq.(\ref{SchwingePairCreation}), the probability of creation of particles in the excited level with mass $M+m$ and charge $q$  can be expressed in terms of the probability of creation of particles in the ground state with mass $M$ and charge $q$ with the extra term:
\begin{eqnarray}
\Gamma^{\rm Schw}(M+m) =\Gamma^{\rm Schw}(M)   \exp{\left(- \frac{2\pi  Mm}{q\cal E} - \frac{\pi  m^2}{q\cal E} \right)}.
\label{Mm}
\end{eqnarray}
Comparing this with Eq.(\ref{SchwingerUnruhDouble}) one obtains that the effect of acceleration (the Unruh effect) is described by the following equation:
 \begin{eqnarray}
   \Gamma_\pm^{\rm Unruh}(m)=   \exp{\left(- \frac{\pi m}{a} \left(1+ \frac{m}{2M}\right)\right)} \,,
 \label{UnruhCoh}
 \end{eqnarray}

In the limit $m\ll M$ this corresponds to Eq.(\ref{SchwingerAcceleration}) with double Unruh temperature.
So, we returned back to the factor 2 problem. But now we got the new argument in favour of the factor 2 in the Unruh effect.  The combined Schwinger-Unruh process suggests that the factor 2 in the equation (\ref{SchwingerAcceleration}) is natural, and thus we must reconsider the temperature of the pure Unruh effect.
 Condensed matter analogs, where we know both the infrared and ultraviolet limits,\cite{Volovik2023u} support doubling. This shows that the double Unruh temperature $\bar T_{\rm U}=2T_{\rm U}$ is not an artefact, but a real temperature that governs the Unruh process.

 \subsection{Back reaction of detector in Unruh effect}
 \label{UnruhSec}

The equation (\ref{UnruhCoh}) suggests that the Unruh effect should be written in the form
\begin{eqnarray}
  \Gamma^{\rm Unruh}(m)=   \exp{\left(- \frac{m}{\bar T_{\rm U}} \left(1+ \frac{m}{2M}\right)\right)} \,,
 \label{UnruhCorrected}
  \end{eqnarray}
where the correction $m/2M$ describes the recoil (or back reaction) of the detector due to its finite mass $M$.\cite{Parentani1997,Reznik1998,Casadio1999,KharzeevTuchin2005} This is similar to the back reaction of the black hole in Hawking radiation discussed by Parikh and Wilczek.\cite{ParikhWilczek2000}

To get this correction let us consider the step by step transitions to the higher mass of the detector, i.e. to the higher level $n$ of excitation of the detector.\cite{Reznik1998,Casadio1999} We consider $N$ steps, each with $\Delta m=m/N$. At each step the mass of the detector increases, $M_n=M+(n-1)\Delta m$ and thus the acceleration and the Unruh
temperature decrease correspondingly:
\begin{eqnarray}
 M_n=M+(n-1)\Delta m \,\,,\,\,  \Delta m=\frac{m}{N}\,,
  \label{steps}
  \\
  a_n=\frac{q{\cal E}}{M_n} \,\,,\,\, \bar T_{{\rm U}n}=\frac{a_n}{2\pi}\,.
 \label{steps2}
 \end{eqnarray}
 Then one obtains Eq.(\ref{UnruhCorrected}):
\begin{eqnarray}
  \Gamma^{\rm Unruh}(m)=\prod_n \exp{\left(- \frac{\Delta m}{\bar T_{{\rm U}n}} \right)}=  \exp{\left(- \sum_{n=1}^N \frac{\Delta m}{\bar T_{{\rm U}n}} \right)}\,,
 \label{Unruh2}
 \\
=   \exp{\left(- \frac{m}{\bar T_{\rm U}} \left(1+ \frac{m}{2M}\right)\right)} \,.
 \label{Unruh3}
  \end{eqnarray}

 \subsection{Back reaction in Hawking radiation}
 \label{BackReactionSec}
 
 The same step-wise procedure can be applied to the Hawking radiation, where the black hole mass $M$ decreases after each $\Delta \omega=\omega/N$ step of Hawking radiation which step by step raises the Hawking temperature.\
 \begin{eqnarray}
 M_n=M-(n-1)\Delta \omega \,\,,\,\,  \Delta \omega=\frac{\omega}{N}\,,
  \label{HawkinhSteps}
  \\
 T_{{\rm H}n}=\frac{1}{8\pi GM_n}\,.
 \label{HawkingSteps2}
 \end{eqnarray}
 This gives the Parikh-Wilczek result:\cite{ParikhWilczek2000}
 \begin{eqnarray}
  \Gamma^{\rm Hawking}(\omega)=
    \exp{\left(- \sum_{n=1}^N \frac{\Delta \omega}{T_{{\rm H}n}} \right)}\,,
 \label{Hawking3}
 \\
=   \exp{\left(- \frac{\omega}{T_{\rm H}} \left(1- \frac{\omega}{2M}\right)\right)} \,\,,\,\, T_{\rm H}=\frac{1}{8\pi GM}\,.
 \label{Hawking4}
  \end{eqnarray}
  The result is similar to that in Eq.(\ref{Unruh3}) except for the opposite sign, since the mass of the black hole decreases with radiation, while the mass of the detector increases due to excitation by acceleration.

 \section{Landau-Khalatnikov hydrodynamics and cosmological constant problems}
 \label{CosmologicalProblem}

\subsection{Vacuum energy in condensed matter}
 \label{VacuumCondSec}

The cosmological constant problem or "vacuum catastrophe" is the substantial disagreement between the observed values of vacuum energy density (the small value of the cosmological constant) and the much larger theoretical value of zero-point energy suggested by quantum field theory. The zero-point energy is huge, because it is determined by the ultraviolet cut-off.
 At first glance the condensed matter has a very simple solution to this problem.
 
In condensed matter systems, such as superfluid $^4$He and superfluid phases of $^3$He, we know physics both at the ultraviolet (atomic) scale and in the infrared (hydrodynamic) limit.  In this low-energy limit, physics is described by collective variables and quasiparticles forming the analogue of the Universe. The effective quantum fields in these effective Universes have zero-point energies. These energies are formally diverging (in the same way as in our Universe) and require the ultraviolet cut-off. 
However, looking from the microscopic (atomic ultraviolet) side we can see that all these divergencies are formal and have no physical consequences. They do not determine the energy of the ground state of the quantum liquid --  the energy of the "quantum vacuum", which is zero in equilibrium. 

The reason is that the "vacuum" is the many-body system. In the limit of large number of atoms it obeys the general laws of thermodynamics, which do not depend on details of the atomic physics. In particular, the energy density $\epsilon$ of the superfluid liquid in the zero temperature limit obeys the Gibbs-Duhem relation:
\begin{equation} 
\epsilon - \mu n=-P \,.
\label{GibbsDuhemHe}
\end{equation}
Here $n$ is the density of atoms, $\mu=d\epsilon/dn$ is the chemical potential and $P$ is pressure.

One can see that the thermodynamic potenial $\epsilon - \mu n$ plays the role of the energy density of the vacuum, 
$\epsilon_{\rm vac}$, which obeys the equation of state $w=-1$:
 \begin{equation} 
\epsilon_{\rm vac}=-P \,\,, \,\,  \epsilon_{\rm vac}=\epsilon - \mu n \,.
\label{rhovac}
\end{equation}

\subsection{Cosmological constant is zero in equilibrium}
 \label{ZeroEquilibriumSec}
 
Eq.(\ref{rhovac}) is the general property of any vacuum state in any quantum system including the quantum vacuum of our Universe. It is important that it is $\epsilon_{\rm vac}$, and not $\epsilon$, that enters into Einstein's equation as the cosmological constant $\Lambda$. 
This is shown using the so-called $q$-theory,\cite{KlinkhamerVolovik2008a,KlinkhamerVolovik2008b,KlinkhamerVolovik2008c} where the physical vacuum is described by the 4-form field -- the dynamical variable introduced by Hawking.\cite{Hawking1984,Duff+Wu} From the equations for $q$-field and Einstein equation one obtains:
 \begin{equation} 
\Lambda=\epsilon_{\rm vac}=\epsilon(q) -\mu q\,\,,\,\, \mu=\frac{d\epsilon}{dq}\,.
\label{Lambda}
\end{equation}

Let us first consider the quantum vacuum in the absence of gravity, i.e. in special relativity.\cite{Volovik2003c}
In this case if the external pressure acting on the quantum vacuum is
absent, then according to Eq.(\ref{rhovac}), the cosmological constant is naturally zero. 
So, the thermodynamics naturally solves the main cosmological constant problem without any fine-tuning, if gravity is absent.
Note that the quantity $\epsilon(q)$ is determined by the UV physics, and is huge. But in equilibrium it is naturally cancelled by the term $-\mu q$, which is the analogue of the counterterm in quantum field theories. The mechanism of cancellation is purely thermodynamic and does not depend on whether the vacuum is relativistic or not. 

Hawking considered the quadratic dependence of $\epsilon$ on $q$.
In this case there is no compensation, since one obtains $\Lambda=\epsilon_{\rm vac} = -\epsilon(q)$.
In the analogy with condensed matter, this vacuum corresponds to a gas instead of a liquid. A gas cannot exist in equilibrium without external pressure, whereas a liquid is self-sustained, i.e. can exist without external pressure. 
This suggests that the vacuum in our Universe is the self-sustained substance, and thus  $\Lambda=\epsilon_{\rm vac}=0$ in equilibrium.

\subsection{Vacuum decay without fine-tuning}
 \label{FineTuningSec}
 
Now consider the influence of gravitational degrees of freedom. As shown in section \ref{dSthermodynamicsSec}, gravity with its thermodynamically conjugate variables (the gravitational coupling $K$  and the scalar Riemann curvature ${\cal R}$) adds its contribution to the total pressure. Then at equilibrium the total pressure is zero according to eq. (\ref{Equilibrium3}). But now the equilibrium state with zero pressure corresponds to the de Sitter vacuum, which by virtue of its symmetry serves as an attractor along with the Minkowski vacuum.

The de Sitter attractor can be considered using the dynamics of the Hawking $q$-field.  Let us first assume that the Big Bang has started in the originally equilibrium Minkowski vacuum. This corresponds to the special value $\mu_0$ of the chemical potential of the $q$ field in Eq.(\ref{rhovac}) for which $\Lambda(\mu_0)=0$. Then from our equations without dissipation\cite{KlinkhamerVolovik2008b}  it follows that 
the cosmological constant, which is very large immediately after the Big Bang, relaxes with oscillations to zero value. Its magnitude averaged over fast oscillations reaches the present value in the present time $\left<\Lambda(t_{\rm present})\right> \sim M^2_{\rm P}/t^2_{\rm present} \sim 10^{-120} M^4_{\rm P}$.
The similar behaviour was discussed in Sec. \ref{EnergyExchangeSec}.
The oscillating decay also present in the Starobinsky inflation,\cite{StarobinskyYokoyama1994,Kofman1997} where the magnitude of the oscillation frequency is determined by the Higgs inflaton mass, instead of the Planck scale in our case.

However, if the initial conditions are different, i.e. $\mu \neq \mu_0$, then from the equations of dynamics (again without dissipation) it follows that the Universe relaxes with oscillations to de Sitter attractor instead of the Minkowski vacuum state. This demonstrates that the special choice of initial conditions, $\mu=\mu_0$, is a kind of fine-tuning. The fine-tuning is also present in the Starobinsky inflation,\cite{StarobinskyYokoyama1994,Kofman1997} due to the special choice of the parameters of system, which leads to the Minkowski attractor. The Pauli-Zeldovich mechanism discussed by Kamenshchik and Starobinsky \cite{KamenshchikStarobinsky2018} also relies  on the fine-tuning -- the exact cancellation of contributions of relativistic bosons and relativistic fermions to the vacuum energy.
In general case, i.e. without fine-tuning, the de Sitter attractor is inevitable. This reflects a special symmetry of the de Sitter state that puts this state on the same level as the Minkowski vacuum. 

 Nevertheless, the condensed matter conclusion that the vacuum energy must be zero at full equilibrium still holds. The de Sitter vacuum is unstable and decays to the Minkowski vacuum. This happens because the de Sitter environment serves as the heat bath for matter. The thermal bath provides the thermal nucleation of matter discussed in Section \ref{HeatBathSec}. The created matter violates the peculiar symmetry of the de Sitter state. The energy exchange between vacuum and matter finally leads to the decay of this state and thus to the nullification of the cosmological constant in the final thermodynamic equilibrium without any fine-tuning. This was considered in Sections \ref{dSthermodynamicsSec} and \ref{dSdynamicsSec}. The  relaxation of the dark energy Eq.(\ref{DecayLaw}) was obtained on example of the energy exchange with the Zel'dovich stiff matter.

This mechanism of instability of the de Sitter vacuum towards the Minkowski vacuum differs from the instability considered by Polyakov. \cite{Polyakov2008,Polyakov2010,KrotovPolyakov2011,Polyakov2018} In our case the radiation of matter in the de Sitter environment takes place only if the de Sitter state contains an "impurity", such as proton or hydrogen atom in Section \ref{HeatBathSec}. This impurity breaks de Sitter symmetry and becomes the trigger for the creation of matter.
 
Also note that the decay of the "cosmological constant" cannot be obtained in the frame of the classical Einstein equations due to the Weinberg no-go theorem.\cite{Weinberg1989} This process requires the quantum mechanical approach, which includes the quantum tunneling.
In condensed matter, this can be compared with the classical time-dependent Ginzburg-Landau equation for superfluids and superconductors. It has a very narrow range of applicability, and nonequilibrium processes such as relaxation to an equilibrium state require the Green's function approach, see Kopnin's book.\cite{KopninBook}
 
\section{Planck constants and dimensionless physics}
\label{PlanckConstantsSec}

\subsection{Composite tetrads from relative symmetry breaking}
\label{CompositeTetradSec}

There are various scenarios for the emergence of gravity from more fundamental fields, such as quantum fermionic fields. In particular, gravity can be constructed using the bilinear combinations of the fermionic fields:
\cite{Akama1978,Diakonov2011,VladimirovDiakonov2012,VladimirovDiakonov2014,ObukhovHehl2012,Maiezza2022,Maitiniyazi2025}
\begin{equation}
 \hat E^a_\mu = \frac{1}{2}\left( \Psi^\dagger \gamma^a\partial_\mu  \Psi -  \Psi^\dagger\overleftarrow{\partial_\mu}  \gamma^a\Psi\right) \,.
\label{TetradsFermionsOperators}
\end{equation}
The original action in this scenario does not depend on the gravitational fields (tetrads and metric) and is described solely in terms of differential forms:
\begin{equation}
S=\frac{1}{24}e^{\alpha\beta\mu\nu} e_{abcd} \int d^4x  \, \hat E^a_\alpha   \hat E^b_\beta \hat E^c_\mu \hat E^d_\nu \,.
\label{OriginalActions}
\end{equation}
This action, which is the operator analog of the cosmological term, has high symmetry. It is symmetric under coordinate transformations $x^\mu \rightarrow \tilde x^\mu(x)$, and thus is also scale invariant. In addition, the action is symmetric under spin rotations, or under the corresponding gauge transformations when the spin connection is added to the gradients. 

The gravitational tetrads $e^a_\mu$ appear as the vacuum expectation values of the bilinear fermionic 1-form $\hat E^a_\mu$ as a result of the spontaneous symmetry breaking:
\begin{equation}
e^a_\mu=<\hat E^a_\mu>\,.
\label{TetradsFermions}
\end{equation}
This order parameter breaks the separate symmetries under orbital and spin transformations, but remains invariant under the combined rotations. On the level of the Lorentz symmetries the symmetry breaking scheme is $L_L \times L_S \rightarrow L_J$. Here $L_L $ is the group of Lorentz transformations in the coordinates space, $L_S $ is the group of Lorentz transformations in the spin space, and 
$L_J$ is the symmetry group of the order parameter $e^a_\mu$, which is invariant under the combined Lorentz transformations $L_J$. 

Similar symmetry breaking mechanism of emergent gravity  is known in condensed matter physics, where the effective gravitational vielbein also emerges as the bilinear fermionic 1-form.\cite{Volovik1990}
This scenario takes place in the $p$-wave spin-triplet superfluid $^3$He-B, where the corresponding relative symmetry breaking \cite{Leggett1973} occurs between the spin and orbital rotations, $SO(3)_L \times SO(3)_S \rightarrow SO(3)_J$. This means that the  symmetry under the relative rotations in spin and orbital spaces is broken, while the properties of $^3$He-B are invariant under combined rotations $SO(3)_J$ and thus remain isotropic.

\subsection{Dimensionful metric and dimensionless interval}
\label{IntervalSec}

In the tetrad gravity, the metric field is the bilinear combination of the tetrad fields:
 \begin{equation}
g_{\mu\nu}=\eta_{ab}e^a_\mu e^b_\nu \,,
\label{DimMetric}
\end{equation}
and thus in this scenario of quantum gravity the metric is the fermionic quartet. In principle, the signature of $\eta_{ab}$ can also come from the dynamical variable $O_{ab}$,\cite{BondarenkoZubkov2022,Bondarenko2022} if $\eta_{ab}$ emerges as the vacuum expectation value of the corresponding symmetry breaking phase transition, $\eta_{ab}=<O_{ab}>$.

Ii is important that in this quantum gravity, the fermionic fields $\Psi$ are dimensionless\cite{VladimirovDiakonov2012} (we denote this as $[\Psi]=[1]$, i.e. $\Psi$  has the same zero dimension as numbers). Thus the tetrads in Eq.(\ref{TetradsFermions}) have the dimensions of the inverse time and inverse length, $[e^a_0]=1/[t]$ and
 $[e^a_i]=1/[L]$, while the metric elements in Eq.(\ref{DimMetric}) have dimensions $1/[t]^2$, $1/[L]^2$
 and $1/[t][L]$.
 Due to such dimensions of tetrads and metric, the interval  is dimensionless:
  \begin{equation}
ds^2=g_{\mu\nu}dx^\mu dx^\nu \,\,, \,\,  [s^2]=[1] \,.
\label{DimensionInterval}
\end{equation}
This is not surprising: the interval is a diffeomorphism invariant, whereas in this approach to quantum gravity all diffeomorphism-invariant quantities (such as action (\ref{OriginalActions})) are dimensionless.\cite{VladimirovDiakonov2012}

\subsection{Dimensions of gauge field and mass} 
\label{GaugeFieldSec}

Let us consider the simplest examples of the dimensionless action.  The action describing interaction of a charged point particle with the $U(1)$ gauge field is:
 \begin{equation}
S=q \int dx^\mu A_\mu \,.
\label{ChargeParticleAction}
\end{equation}
 As the original action (\ref{OriginalActions}), this action does not depend on the metric field and is described solely in terms of differential forms, now in terms of 
 the 1-form $U(1)$ gauge field $A_\mu$. 
 
 The $U(1)$ field is the geometric quantity, which comes from the gauging of the global $U(1)$ field.  The field $A_\mu$ comes from the gauging of gradient of the phase field, and thus has dimension of the gradient of phase, with $[A_0]=1/[t]$ and $[A_i]=1/[L]$.
The charge  $q$ here is dimensionless -- it is the integer (or fractional) geometric charge of the fermionic or bosonic field.  In case of electromagnetic $U(1)$-field, $q$ is expressed in terms of the electric charge of electron, i.e. $q=-1$ for electron and $q=+1$ for proton. As a result the action (\ref{ChargeParticleAction}) is naturally dimensionless, $[S]=[1]$.

Such action can be extended to the objects of higher dimensions, which interact with the corresponding gauge fields:  $1+1$ strings interacting with 2-form gauge field, $2+1$ branes interacting with the 3-form field and also $3+1$ objects interacting with the 4-form field.

The action describing the classical dynamics of a point particle requires  the metric field, since it is expressed in terms of the  interval:
 \begin{equation}
S=M\int ds  \,.
\label{particleAction}
\end{equation}
Since both the interval $ds$ and the action $S$ are  dimensionless, from equation (\ref{particleAction})  it  follows that the particle mass  $M$ is also dimensionless, $[M]=[S]=[s]=[L]^0=[t]^0=[1]$. 
Note that we discuss here the space and time dimensions of the physical quantities. In quantum field theories, the mass dimension is also used, which will be discussed in Sec. \ref{PlanckDimensions}.

\subsection{Two Planck constants}
\label{TwoPlanckSec}

Let us consider the quadratic terms in the action for the classical scalar field $\Phi$:
\begin{equation}
S=\int d^4 x\,\sqrt{-g} \,\left(  g^{\mu\nu} \nabla_\mu \Phi^*  \nabla_\nu \Phi +M^2|\Phi|^2 \right)\,.
\label{scalar}
\end{equation}
Comparing the gradient term and the mass term, and using the dimension of the metric, one again obtains that the mass $M$ is dimensionless, $[M]=[1]$.
Then since the action $S$ and volume element  $d^4 x\,\sqrt{-g}$ are dimensionless, it follows that the scalar field is also dimensionless,  $[\Phi]^2=[M]=[S]=[1]$.

Using the Fourier components, one obtains the spectrum of waves:
\begin{equation}
g^{\mu\nu} k_\mu k_\nu + M^2=0\,,
\label{FourierSpectrum}
\end{equation}
where $k_0=\omega$ is the frequency of the mode and ${\bf k}$ is the wave vector.
In Minkowski spacetime one obtains
\begin{equation}
E^2 = M^2+{\bf p}^2c^2 \,,
\label{FourierESpectrum}
\end{equation}
where energy and momentum of particles are 
\begin{eqnarray}
E=\sqrt{-g^{00}}\,\omega \,,
\label{Energy}
\\
 cp_x =\sqrt{g^{xx}}\,k_x\,\,\,,\,\, cp_y =\sqrt{g^{yy}}\,k_y \,\,,\,\, cp_z =\sqrt{g^{zz}}\,k_z \,.
\label{Momentum}
\end{eqnarray}
This shows that the Minkowski elements of metric play the role of the Planck constants $\hbar$ and $\nh\equiv\hbar c$, which connect the energy and momentum of particles with the frequency and the wave vector. This suggests, that we have the following relations between the Planck constants and  the Minkowski metric:
\begin{equation}
-g^{00}_{\rm Mink} \equiv\hbar^2\,\,,\,\, g^{ik}_{\rm Mink}\equiv\nh^2 \delta^{ik}\,.
\label{MinkowskiMetric1}
\end{equation}

Both Planck constants, bar $\hbar$ and slash 
$\nh$, are the elements of the metric and tetrads in Minkowski vacuum. The bar $\hbar$ is the time component of the inverse tetrad and has dimension of time, $[\hbar]=[t]$, while the slash 
$\nh$ enters the space components of tetrads and has dimension of length, $[\nh]=[L]$.  The presence of the Planck constants in the metric elements demonstrates that the metric field describes the dynamics of the quantum vacuum, rather than the space-time geometry. This is natural, since space-time itself with its metric arises from the dynamics of quantum fields. The speed of light, as an element of geometry, arises as the ratio of two Planck constants:
\begin{equation}
 c^2=\frac{\nh^2}{\hbar^2}\,.
\label{SpeedOfLight}
\end{equation}

In principle, there can be more than two Planck constants. For example,
two Planck constants $\hbar$ and $\hbar_g$ were introduced in Ref. \cite{Terno2024}
 for matter and gravity correspondingly (see also \cite{Caro1999}). This suggests that gravity will be more classical than matter if $\hbar_g\ll \hbar$. With slash component $\nh_g$ added one obtains four different Planck constants: $\hbar$, $\nh$, $\hbar_g$ and slash $\nh_g$. Then in principle one may have two speeds of light: $c=\nh/\hbar$ and $c_g=\nh_g/\hbar_g$, although $c_g=c$ is more natural.

 \subsection{Planck constants do not enter diffeomorphism invariant expressions}
 \label{PlanckDimensions}

The diffeomorphism invariant quantities are dimensionless. This includes the action $S$ (example is in Eq.(\ref{OriginalActions})); the scalar curvature ${\cal R}$; the scalar field $\Phi$; the wave function $\psi$; masses $M$; the cosmological constant $\Lambda$; the gravitational coupling $K$; etc.\cite{Volovik2021,Volovik2021cont}
Here the term "dimensionless" means the absence of space or time dimensions. However, these dimensionless quantities may have nonzero mass dimension $[M]$. For example, action has zero mass dimension; the mass dimension of curvature is $-2$, i.e. $[{\cal R}]=[M]^{-2}$; the mass dimension of gravitational coupling is 2, i.e. $[K]=[M]^2$; the cosmological constant $\Lambda$ has mass dimension 4.

Since $M$ is spacetime dimensionless, from equation $M=\hbar \omega$ it follows that the Planck constant has dimension of time, $[\hbar]=[t]$, and the slash Planck constant has dimension of length, $[\nh]=[L]$. Since these "constants" are not diffeomorphism invariant, they cannot enter the diffeomorphism invariant expressions. Example is the original action (\ref{OriginalActions}), which does not contain the Planck constants $\hbar$ and $\nh$. This is the property of any action, if it is written  in the diffeomorphism-invariant form. Let us consider several examples, when the Planck constants are removed from the expressions rewritten in diffeomorphism-invariant form.

The gravitational potential must be expressed in terms of the rest energies $M=mc^2$ instead of the masses $m$. Thus the Newton constant must be modified:
\begin{eqnarray}
U(r)=-{\bar G}\frac{M_1M_2}{r}\,\,,\,\, {\bar G}=\frac{G}{c^4}\,.
\label{RestMass}
\end{eqnarray}
 Since the potential $U$ has dimension of mass $M$, and thus is the space-time dimensionless, one obtains that  ${\bar G}$ has the same dimension of length as $\nh$, i.e. 
 $[{\bar G}]=[{\nh}]=[L]$. This demonstrates that both ${\bar G}$ and $\nh$  are not diffeomorphism invariant, and thus they cannot be the fundamental constants. In the gravitational action they compensate each other: 
\begin{eqnarray}
S =K \int d^4x\,\sqrt{-g} \,{\cal R} \,\,,\,\, K= \frac{1}{16\pi} \frac{\nh}{{\bar G}} \equiv \frac{M_{\rm P}^2}{2}\,.
\label{EinsteinAction4D}
\end{eqnarray}
As we already mentioned, the gravitational coupling $K$ and curvature ${\cal R}$ are space-time dimensionless, but have mass dimensions 2 and $-2$ correspondingly. That is why, although $\nh$ and  ${\bar G}$ have the same space-time dimension, their ratio has mass dimension 2. The length of the reduced Planck constant $\nh$ and the reduced Planck length $l_{\rm P}=\nh/M_{\rm P}$ coincide if we choose the reduced Planck mass $M_{\rm P}$ as unity of mass.

Another example is the Eq.(\ref{Classical1}) for electron levels in hydrogen atom, which can be written in the conventional form or in terms of the dimensionless quantities. In the traditional form it is
\begin{eqnarray}
E_n= \frac{me^4}{2\hbar^2}\frac{1}{n^2}\,.
\label{EnergyLevelsConv}
\end{eqnarray}
In the dimensionless form these levels can be considered in terms of the negative corrections to the electron rest energy:
\begin{eqnarray}
\frac{\Delta M_e}{M_e}=- \frac{\alpha^2}{2n^2}\,.
\label{EnergyLevelsDimensionless}
\end{eqnarray}
In Eq.(\ref{EnergyLevelsDimensionless}), the fully dimensionless parameter $\alpha$ (the fine structure constant $\alpha=e^2/\nh$) connects the space-time dimensionless quantities:  the electron rest energy $M_e$ and its correction $\Delta M_e$. 

The traditional description in Eq.(\ref{EnergyLevelsConv}) reflects the historical process of development of physical ideas. Here the dimensionless quantities are split into dimensional quantities, such as electric charge $e$, speed of light $c$, Newton constant $G$, Planck constant $\hbar$, mass $m$, Hubble parameter $H$, etc.

 \subsection{de Sitter as ensemble of Planck "atoms"}
 \label{dSensemble}

In principle, some dimensionless parameters can be quantized. The numbers related to symmetry and topological charges are certainly quantized. The other possible example is the entropy of the black hole or the entropy of the Hubble volume, which are related to the event horizons. 

The energy of the Hubble volume is obtained by multiplying the energy density in the equation (\ref{dSEnergyDensity}) by the volume $V_H$.  It can be written in terms of the space-time dimensionless variables, Planck mass $M_{\rm P}$ and de Sitter temperature $T=\hbar H/\pi$:
\begin{equation}
E_H=\epsilon_{\rm vac} V_H= \frac{1}{2\hbar H}\frac{\nh}{\bar G}=\frac{4M_{\rm P}}{T} \,M_{\rm P}\equiv NM_{\rm P}\,.
\label{dSenergy}
\end{equation}
In this equation the energy is presented in terms of $N$ "atoms of de Sitter state" with the Planck masses $M_{\rm P}$.
In this representation, the dimensionless entropy of the Hubble volume in Eq.(\ref{dSEntropyHubble}) is quantized:
\begin{equation}
S_H= \frac{A}{4\nh \bar G}= \frac{\pi}{(\hbar H)^2}\frac{\nh}{\bar G}= 8\frac{M_{\rm P}^2}{T^2} =\frac{1}{2}N^2\,.
\label{dSentropy}
\end{equation}
This leads to quantization of the cosmological constant.\cite{Volovik2025P}

The entropy of the Hubble volume is equivalent to the entropy of the cosmological horizon. Quantization of the horizon entropy is considered in many theories. In the Bekenstein approach, \cite{Bekenstein1974,Mukhanov1986,Kastrup1997,Khriplovich2008,Dvali2011,Kiefer2020,Bagchi2024}
the entropy is the adiabatic invariant, which spectrum is equally spaced giving rise to the linear law, $S_{\rm BH}=a N$. Here $a$ is the dimensionless parameter, which value depends on the microscopic theory. The quantization inspired by the string theory\cite{Gibbons2011,Visser2012} gives the square-root rule, $S =2\pi(\sqrt{N_1} \pm \sqrt{N_2})$
Note that the equation (\ref{dSentropy}) applies to the entropy of the Hubble volume, whereas in general the entropy of the de Sitter state is extensive, see section \ref{dSextensive}.

 \subsection{Black hole as ensemble of Planck "atoms"}
 \label{BHensemble}
 
The ensemble of the $N$ Planck "atoms" in Eq.(\ref{dSenergy}) is similar to the ensemble of $N$ Planck size black holes inside the black hole horizon,\cite{Volovik2025,Volovik2025P} where the entropy is also proportional to $N^2$ at large $N$:
\begin{equation}
S(N)=\frac{N(N-1)}{2}\,.
\label{Nsquare}
\end{equation}
The probability of the quantum tunneling process of the emission of the black hole quantum 
($N=1$ and $S(N=1)=0$) is, on the one hand, similar to Eq.(\ref{VortexNucleation}) for quantum tunneling process of the vortex instanton, and on the other hand, it satisfies the equation (\ref{HoleEmession2}) for quantum tunneling as random fluctuation:
\begin{equation}
w \sim  e^{S(N-1) -S(N) } = e^{-(N-1)} =\exp\left(-\frac{m_{\rm P}}{T_{\rm H}}\left(1-\frac{m_{\rm P}}{M}\right)\right)\,.
\label{QuantumEmission}
\end{equation}
The correction to the Hawking radiation rate corresponds to the effect of back reaction in the process of radiation considered by Parikh and Wilczek.\cite{ParikhWilczek2000}

 The black hole entropy is non-extensive and thus requires the generalized statistics.
For large $N$, the entropy in Eq.(\ref{Nsquare}) satisfies the composition rule of the Tsallis-Cirto 
$\delta=2$ entropy for the ensembles with $N_1$ and $N_2$ quanta:
\begin{equation}
\sqrt{S_{\delta=2} (N_1 + N_2)}= \sqrt{S_{\delta=2}(N_1)}+ \sqrt{S_{\delta=2}(N_2)}\,,
\label{CompositionN}
\end{equation}
which reproduces the Eq(\ref{TwoBlackHoles}). In rare quantum fluctuation processes leading to black hole fragmentation, $N= N_1+N_2$, the entropy decreases sharply. Then the entropy continuously increases in the further process of black hole recombination, $N_1+N_2\rightarrow N$, reaching a maximum value for a given $N$. In this description, a black hole is not a quantum state, but a dissipative process in which continuous evolution alternates with random quantum jumps, see e.g. Refs.\cite{Cohen-Tannoudji1993,Frohlich2024}.

It would be interesting to consider the analogs of such processes and possible non-extensive statistics in superfluids. 
One direction is to consider entropy associated with the vortex instanton in Eq.(\ref{VortexNucleation}), which describes the creation of vortices in the moving superfluid $^4$He. In the toroidal geometry this instanton changes the topological charge of the flow -- the  number of quanta of circulation. Let us assume that equation (\ref{VortexNucleation}) is related to the entropy of the flowing "vacuum" and compare this entropy with the black hole entropy. For that let us introduce the corresponding mass $M$, which is the energy of the created vortex loop, and the corresponding Planck mass $M_{\rm P}$, which is the inverse interatomic distance. Then one obtains that the entropy corresponding to Eq.(\ref{VortexNucleation}) is $S\propto M^3/M_{\rm P}^3$. This entropy has higher non-extensivity than the black hole entropy $S\propto M^2/M_{\rm P}^2$ and it corresponds to the Tsallis-Cirto entropy with $\delta=3$. The corresponding effective temperature, which characterizes the "flow of the vacuum" with velocity $v_s$, is 
$T=dM/dS \propto m_4 v_s^2$, where $m_4$ is the mass of the helium atom.

Another possible direction is to consider the quantum tunneling process of splitting a droplet of superfluid liquid into smaller droplets with a decrease in the total entropy. In particular, due to the surface energy of the droplets, it is possible to split a droplet with a non-zero temperature and, therefore, with non-zero entropy into smaller parts with zero temperature and zero entropy.

 \subsection{Acoustic Planck constants}
 \label{AcousticSec}
 
 Returning to the two-fluid hydrodynamics, let us consider quasiparticles in Bose superfluid $^4$He.
The flow of the liquid provides the acoustic gravity for "relativistic" quasiparticles -- phonons. The effective acoustic metric and thus the acoustic Planck constants, $\hbar_{\rm ac}$ and $\nh_{\rm ac}$, can be expressed in terms of the parameters of this liquid. 

If the chosen variable in the superfluid hydrodynamics is the phase $\Phi$ of the Bose condensate, which is dimensionless, then the action for phonons propagating in moving liquid is:\cite{Volovik2003,Volovik2023c}
\begin{eqnarray}
S_{\rm ph}=\frac{m_4}{2\hbar} \int d^3x dt \, n \left( (\nabla \Phi)^2  - \frac{1}{s^2} (\dot \Phi -{\bf v} \cdot \nabla \Phi)^2\right)=\,
\nonumber
\\
= \frac{1}{2}\int d^4 x\sqrt{-g} \, g^{\mu\nu} \nabla_\mu \Phi \nabla_\nu \Phi \,.\,\,\,\,
\label{PhononAction}
\end{eqnarray}
Here $n=1/a^3$ is the density of bosonic particles ($^4$He atoms, which play the role of the "atoms of the vacuum"); $a$ is the interatomic distance, which plays the role of the Planck length; $m_4$ is the mass of the helium atom; $s$ is the speed of sound, which plays the role of the speed of light; ${\bf v}$ is the superfluid velocity (the velocity of the "superfluid vacuum"). It is the shift function in the Arnowitt-Deser-Misner (ADM)  approach. 

Finally, $\hbar$ here is the conventional Planck constant describing the microscopic physics. All the parameters are considered in the conventional units, i.e. without application of dimensionless physics. The dimensionless physics will emerge for phonons, and this is the reason why we introduced the phonon action $S_{\rm ph}$ as the action divided by $\hbar$.

The corresponding acoustic interval is
\begin{equation}
ds^2=g_{\mu\nu}dx^\mu dx^\nu=\frac{\hbar n}{m_4s}[-s^2 dt^2 + (d{\bf r} -{\bf v}dt)^2]\,.
\label{AcousticInterval}
\end{equation}
The analog of the Minkowski metric corresponds to the zero value of the shift function, ${\bf v}=0$, and thus $g^{0i}=0$. Then
the effective acoustic Minkowski metric experienced by the propagating phonons is:
\begin{eqnarray}
g_{00}=\frac{\hbar n s}{m_4}   \,\,,\,\, g_{ik}=\frac{\hbar n}{m_4s}  \delta_{ik} \,\,,\,\,
\sqrt{-g}= \frac{\hbar^2n^2}{m^2s} \,,
\label{Effective_metric}
\end{eqnarray}
with dimensions 
\begin{eqnarray}
[g_{00}]=\frac{1}{[t]^2}   \,\,,\,\, [g_{ik}]=\frac{1}{[L]^2} \,\,,\,\,
[\sqrt{-g}]= \frac{1}{[t][L]^3}  \,.
\label{EffectiveMetricDimension}
\end{eqnarray}
The acoustic interval (\ref{AcousticInterval}) is dimensionless, $[ds]=1$. This demonstrates that  the interval $ds$ describes the dynamics of phonons in the superfluid "vacuum", rather than the distances and time intervals.
The same is valid for the interval in general relativity, where it describes the dynamics of a point particle in the relativistic quantum vacuum.

The quantum mechanics of phonons demonstrates that it can be described by the effective acoustic Planck constants, $\hbar_{\rm ac}$ and $\nh_{\rm ac}$:
\begin{eqnarray}
\hbar_{\rm ac}^2=g^{00}=\frac{m_4}{\hbar n s}  \,\,,\,\, \nh_{\rm ac}^2=\hbar_{\rm ac}^2s^2 =\frac{m_4s}{\hbar n}  \,.
\label{Effective_hbar}
\end{eqnarray}
The ground state of superfluid $^4$He represents the strongly interacting and strongly correlated "quantum vacuum". Two Planck mass scales $m_4s^2$ and $\hbar s/a$ are of the same order. That is why in this analogue of quantum vacuum the acoustic Planck constant $\nh_{\rm ac} \sim a$, i.e. the length of the effective Planck constant is on the order of the effective Planck length (the interatomic distance).

 \subsection{Phase boundary between Universes with different $\hbar$}
 \label{ContactSec}
 
 The so-called multiple point principle\cite{Nielsen1994} suggests that the quantum vacua are degenerate, and thus there is the boundary of the first order phase transition between our Universe and the Universe with very different properties.\cite{Nielsen2023} If so, it is not excluded that the two neighbouring Universes may have different Minkowski metric and thus different values of Planck parameters $\hbar$ and $\nh$. If there is the thermal contact between these Universes, then in the thermal equilibrium one would have the following condition for temperatures and Planck constants:
\begin{equation}
\frac{T_1}{\hbar_1}=\frac{T_2}{\hbar_2}\,.
\label{TolmanCondition}
\end{equation}
According to Eq.(\ref{MinkowskiMetric1}), which connects the Minkowski metric and Planck constants, the equation (\ref{TolmanCondition}) is the particular case of the Tolman-Ehrenfest relation, 
$T({\bf r}) \sqrt{g_{00}({\bf r})}={\rm constant}$. 

If there is the particle exchange between these two vacua, one has the similar Tolman law for the chemical potentials:
\begin{equation}
\frac{\mu_1}{\hbar_1}=\frac{\mu_2}{\hbar_2}\,.
\label{TolmanChemical}
\end{equation}
Note that the chemical potential is the space-time dimensionless, while $\hbar$ has dimension of time. Then the quantity  $\mu/\hbar$ has dimension of frequency. This means that at equilibrium, when crossing a phase boundary, it is the frequency that does not change, but not the temperature or chemical potential. If the two vacua are superfluid, then the ac Josephson effect between these vacua is determined by the frequency difference:
\begin{equation}
\omega=\bigg | \frac{\mu_1}{\hbar_1}-\frac{\mu_2}{\hbar_2} \bigg |\,.
\label{JosephsonEff}
\end{equation}
For $\hbar_1=\hbar_2$, this equation describes the conventional ac Josephson effect in superfluids.

Another interesting question is what happens when $e_\mu^a \rightarrow 0$, i.e. we approach the critical  point  at which the interval $ds$ becomes identically zero and thus the spacetime disappears. According to Eq.(\ref{MinkowskiMetric1}, in this limit the 
both Planck constants approach infinity, $\hbar\rightarrow \infty$ and $\nh\rightarrow \infty$. This demonstrates the quantum correlations between the points separated by arbitrary large distances. However, there may be different limiting cases corresponding to different relations between $\hbar$, $\nh$ and $G$ in the Bronstein $cGh$ cube,\cite{Bronstein1933,Duff2002} see Ref. \cite{Ecker2025} and references therein.

\section{Discussion}
\label{DiscussionSec}
 
It is shown that the de Sitter Universe can be represented in terms of two dark components in equations (\ref{Equilibrium1})-(\ref{Equilibrium4}). These are the dark energy (the energy of the vacuum) and the "gravitational dark matter", which comes from the gravitational degrees of freedom. The partial pressures of two components compensate each other in the equilibrium de Sitter state.
In this representation, the gravitational dark matter is responsible for the thermodynamics of the de Sitter state, which is characterized by the local temperature $T=H/\pi$ and is similar to the thermodynamics of the Zel'dovich stiff matter. Let us recall that the local temperature of the de Sitter state is twice the Gibbons-Hawking temperature related to the cosmological horizon, and the stiff matter behaviour of the dark matter component is the consequence of this value of the local temperature.

We tried to extend this two-fluid approach to the dynamics of the Universe, assuming that in dynamics the dark matter component is also equivalent to the Zel'dovich stiff matter. Then we obtained the power-law decay of both dark components, the dark matter component in Eq.(\ref{DecayLawMatter}) and the dark energy component in Eq.(\ref{DecayLaw}). While the state immediately after the Big Bang may have an energy density of the order of the Planck scale, $\epsilon \sim M^4_{\rm P}$, the energies of both components acquire the correct order of magnitude at the present time, $\epsilon \sim 10^{-120} M^4_{\rm P}$, thus offering a possible route to solution of the coincidence problem. Both components eventually decay to zero at $t\rightarrow \infty$ on the way to a full equilibrium attractor, the Minkowski quantum vacuum.

The reasonable decay law (\ref{DecayLaw}) for dark energy is obtained in the model with the stiff matter. The question then arises: what might be the physical origin of this stiff matter and how is it related to the gravitational hard matter discussed in the Section \ref{dSthermodynamicsSec}. In principle, it is possible, that the stiff matter comes just from the gravitational dark matter, and the modified equations reflect the microscopic physics, which supports both the thermodynamics and dynamics of the de Sitter. The question then remains open as to how to describe the corresponding degrees of freedom that lead to the considered thermodynamics and dynamics.

One possibility is to  introduce the proper vacuum variable, which describes the quantum vacuum and its vacuum energy density. 
The working example of such variable  is the Hawking 4-form field.\cite{Hawking1984} This variable was used in particular in the so-called $q$-theory.\cite{KlinkhamerVolovik2008c,KlinkhamerVolovik2008G} However, the suggested form of the $q$-theory gives rise to the cold dark matter instead of the stiff matter.\cite{KlinkhamerVolovik2017} The analogue of the cold matter is produced by the fast oscillations of the $q$-field. This $q$-theory mechanism also gives the required power-law decay of the dark energy similar to that in Eq.(\ref{DecayLaw}). However, this decay to the Minkowski vacuum takes place only in case of the proper choice of the value of corresponding analog of the chemical potential. This represents some form of fine tuning, while the two fluid dynamics does not require the fine tuning. 

To avoid fine-tuning the interaction with the ordinary matter must be included. Since
the de Sitter vacuum represents the heat bath for matter, the de Sitter expansion is unstable towards the thermal radiation of matter. Matter violates the de Sitter symmetry and promotes the relaxation of the de Sitter state to the Minkowski vacuum. This is the analogue of the three-fluid hydrodynamics, which describes dark energy, dark matter and ordinary matter.

We also considered the thermodynamics of a black hole, which can be obtained using macroscopic quantum tunneling processes of splitting a black hole into smaller ones. In fact, this is another way to derive the Bekenstein-Hawking entropy of a black hole. It is shown that the statistical ensemble describing a black hole is described by the non-extensive Tsallis-Csirto entropy with $\delta=2$. 

On the other hand the entropy of the de Sitter state is extensive, i.e. for arbitrary volume $V$ in the de Sitter Universe, the entropy is proportional to the volume, $S(V)=sV$, where $s$ is the entropy density. But the integration of the entropy density over the Hubble volume demonstrates the holographic bulk-surface connection between the entropy of the Hubble volume and the Gibbons-Hawking surface entropy of the cosmological horizon, $S(V_H)=sV_H=A/4G$.

\end{document}